\documentclass[a4paper,11pt]{article}
\usepackage{jheppub}

\usepackage{epsfig}
\usepackage{amssymb,amsmath}
\usepackage{pstricks}
\usepackage{pst-all}
\usepackage[utf8]{inputenc}
\usepackage{epstopdf}

\newcommand{\eqa}{\begin{eqnarray}}
\newcommand{\qea}{\end{eqnarray}}
\newcommand{\eq}{\begin{equation}}
\newcommand{\qe}{\end{equation}}

\newcommand{\de}{\mathrm{d}\hspace{0.08em}}

\newcommand{\tr}{\mathrm{Tr}}
\newcommand{\hb}{{\bar{h}}}
\newcommand{\hti}{\tilde{h}}
\newcommand{\Real}{{\mathrm{Re}}}

\newcommand{\avg}[1]{\langle #1 \rangle}

\def\lsi{\raise0.3ex\hbox{$<$\kern-0.75em\raise-1.1ex\hbox{$\sim$}}}
\def\gsi{\raise0.3ex\hbox{$>$\kern-0.75em\raise-1.1ex\hbox{$\sim$}}}
\newcommand{\lsim}{\mathop{\lsi}}
\newcommand{\gsim}{\mathop{\gsi}}

\title{The QCD deconfinement transition for heavy quarks and all baryon chemical potentials}
\author{Michael Fromm,}
\author{Jens Langelage,}
\author{Stefano Lottini,}
\author{Owe Philipsen}
\affiliation{Institut f\"ur Theoretische Physik, Goethe-Universit\"at Frankfurt,\\
Max-von-Laue-Str.~1, 60438 Frankfurt am Main, Germany}
\emailAdd{fromm, langelage, lottini, philipsen@th.physik.uni-frankfurt.de}
\abstract{Using combined strong coupling and hopping parameter expansions, we derive an effective
three-dimensional theory from thermal lattice QCD with heavy Wilson quarks. The theory depends
on traced Polyakov loops only and correctly reflects the centre symmetry of the pure gauge sector
as well as its breaking by finite mass quarks. It is valid
up to certain orders in the lattice gauge coupling and hopping parameter, which can be systematically
improved. To its current order it is controlled for lattices up to $N_\tau\sim 6$ at finite temperature. 
For nonzero quark chemical potentials, the effective theory has a fermionic sign problem which is mild 
enough to carry out simulations up to large chemical potentials. 
Moreover, by going to a flux representation of the partition function, the sign problem can be solved.
As an application, we determine the deconfinement transition and its critical end point as a function of quark mass and
all chemical potentials.}

\begin{document}
\maketitle

\section{Introduction}

The determination of the QCD phase diagram is a fully non-perturbative problem, because QCD is
strongly coupled on the scales relevant to heavy ion collisions and astrophysics, i.e.~for temperatures
$T\lsim 400$ MeV  and baryon chemical potentials $\mu_B\sim 0-3$ GeV. 
On the other hand, a direct first principles approach by Monte Carlo simulations of lattice QCD is 
ruled out by the sign problem, with a complex fermion determinant for quark chemical potential $\mu = \mu_B/3 \neq 0$ prohibiting
importance sampling. Existing workarounds based on reweighting, Taylor expansions in $\mu/T$
or simulations at imaginary chemical potential followed by analytic continuation all introduce 
additional systematic errors and require $\mu/T\lsim 1$ in order to be valid. For an elementary
introduction, see~\cite{oprev}. 
As a consequence, the QCD phase diagram remains largely unknown.

This situation warrants additional investigations by effective theory methods. 
Popular approaches
are based on models which share the chiral and/or the $Z(3)$-symmetry with QCD, such
as PNJL models, Polyakov loop + quark meson models, sigma models etc.
For recent discussions and references, see \cite{Fukushima:2009zz, modrev1,modrev2}.
Other approaches start from QCD directly and employ  Dyson-Schwinger \cite{cf} or functional renormalisation group methods \cite{paw}, using particular truncations. 
A general difficulty with effective theories is to assess the associated
systematic errors. 

Recently, a systematic derivation of a 3d centre-symmetric effective theory for finite temperature
$SU(N_c)$ Yang-Mills by means of a strong 
coupling expansion has been presented, followed by numerical simulations~\cite{Langelage:2010yr}. The effective theory depends on Polyakov loop variables
only, and can be improved order by order.
Its couplings $\lambda_i=\lambda_i(\beta,N_\tau)$ are calculable functions of the lattice gauge coupling 
$\beta$ and the temporal lattice extent
$N_\tau$ of the original
finite temperature lattice theory. The influence of the various couplings can be checked.
Remarkably, the effective theory with only one coupling reproduces the correct order
of the deconfinement transition for $SU(2), SU(3)$, 
and moreover permits a quantitative estimate of the critical couplings $\beta_c(N_\tau)$
for sufficiently fine lattices, such that a continuum extrapolation of $T_c$ is feasible.

In this work we extend the effective theory to include heavy but dynamical fermions of mass $M$ by means of a hopping parameter expansion. This theory permits to explore the phase diagram of QCD with heavy quarks in the $(M,T,\mu)$ parameter space. 
A similar approach to the fermionic sector was taken in 
\cite{DePietri:2007ak}, which however left the gauge sector in the original 4d form.
Here we extend the effective fermionic action 
to order $\kappa^6$ in the hopping expansion. 
As we shall see, for $\mu=0$ we once again obtain good
agreement with full 4d simulations where such results exist. However, the 3d effective theory allows 
for a solution to the sign problem and to explore the full range of quark chemical potentials with 
numerical ease, making 
contact with the region of asymptotically large chemical potentials. 
As a first application of the theory, we map out
the entire deconfinement critical surface delimiting the region of first-order deconfinement 
transitions in the $(M_{u,d},M_s,\mu)$ parameter space.

In Section \ref{sec:efft} we derive the effective theory and discuss its numerical
evaluation, comparing a direct simulation with Polyakov loop degrees of freedom with that of a flux 
representation free of the sign problem. 
Section \ref{sec:res} is devoted to the study of the QCD deconfinement transition, followed by our conclusions and a discussion of further prospects in Section \ref{sec:con}.
\begin{figure}
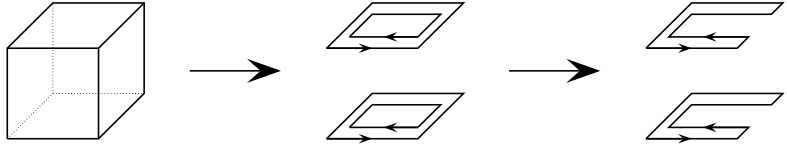

\vspace{1cm}
\hspace{2cm}
\scalebox{0.3}{
\psline[linewidth=2pt,arrowsize=1.0 1.0]{->}(8,3)(12,3)
\psline[linewidth=2pt,arrowsize=0.4 0.4]{->}(14,0)(16,0)
\psline[linewidth=2pt,arrowsize=0.4 0.4]{->}(18,0.5)(16.5,0.5)
\psline[linewidth=2pt,arrowsize=0.4 0.4]{->}(14,4)(16,4)
\psline[linewidth=2pt,arrowsize=0.4 0.4]{->}(18,4.5)(16.5,4.5)
\psline[linewidth=2pt,arrowsize=1.0 1.0]{->}(22,3)(26,3)
\psline[linewidth=2pt,arrowsize=0.4 0.4]{->}(28,0)(30,0)
\psline[linewidth=2pt,arrowsize=0.4 0.4]{->}(32.5,0.5)(30.5,0.5)
\psline[linewidth=2pt,arrowsize=0.4 0.4]{->}(28,4)(30,4)
\psline[linewidth=2pt,arrowsize=0.4 0.4]{->}(32.5,4.5)(30.5,4.5)
\psline[linestyle=dotted](0,0)(2,2)(2,6)
\psline[linestyle=dotted](2,2)(6,2)
\psline[linewidth=2pt](0,0)(4,0)(6,2)(6,6)(2,6)(0,4)(0,0)
\psline[linewidth=2pt](4,0)(4,4)(0,4)
\psline[linewidth=2pt](4,4)(6,6)
\psline[linewidth=2pt](14,0)(18,0)(20,2)(16,2)(14,0)
\psline[linewidth=2pt](14,4)(18,4)(20,6)(16,6)(14,4)
\psline[linewidth=2pt](18,0.5)(19,1.5)(16,1.5)(15,0.5)
\psline[linewidth=2pt](18,0.5)(15,0.5)
\psline[linewidth=2pt](18,4.5)(19,5.5)(16,5.5)(15,4.5)
\psline[linewidth=2pt](18,4.5)(15,4.5)
\psline[linewidth=2pt](28,0)(32,0)(32.5,0.5)(29,0.5)(30,1.5)(33.5,1.5)(34,2)(30,2)(28,0)
\psline[linewidth=2pt](28,4)(32,4)(32.5,4.5)(29,4.5)(30,5.5)(33.5,5.5)(34,6)(30,6)(28,4)
}

\caption{Integration of the cube consisting of 6 plaquettes. In the middle, 
four spatial link variables have been integrated over and we end up with 
two doubly occupied plaquettes. Integrating two of the remaining $U_i$ 
yields the structure on the right which gives a factor 
$[\mathrm{Tr}(U^\dagger U)]^2=N_c^2$.}
\label{fig_cube}
\end{figure}

\section{The effective action}
\label{sec:efft}

\subsection{Finite temperature $SU(N_c)$ Yang-Mills theory}
\label{subsec:YM}

For  the paper to be self-contained and to fix the notation, 
we briefly summarise the main formulae for the $SU(N_c)$
pure gauge case \cite{Langelage:2010yr}. Starting point is the partition function
for the 4d Euclidean Yang-Mills theory with Wilson gauge action,
\begin{eqnarray}
Z&=&\int\left[dU_0\right]\left[dU_i\right]\exp\left[-S_g\right]\;,\nonumber\\
-S_g&=&\frac{\beta}{2N_c}\sum_p
\left(\mathrm{Tr}\;U_p+\mathrm{Tr}\;U_p^{\dagger}\right)\;,\qquad \beta=\frac{2N_c}{g^2}\;.
\label{eq:original_gaugetheory}
\end{eqnarray}
Finite temperature and the bosonic degrees of freedom imply the use of 
periodic boundary conditions in the compactified time direction with $N_\tau$ slices, setting 
the temperature scale by the lattice spacing $a$, $T=1/(aN_\tau)$.
An effective three-dimensional theory emerges by integrating
out the spatial degrees of freedom to get schematically
\begin{eqnarray}
Z&=&\int\left[dU_0\right]\exp\left[-S_\mathrm{eff}\right]\;,\nonumber\\
-S_\mathrm{eff}&=&\ln\int\left[dU_i\right]\exp\left[\frac{\beta}{2N_c}\sum_p
\left(\mathrm{Tr}\;U_p+\mathrm{Tr}\;U_p^{\dagger}\right)\right]
\equiv\lambda_1S_1+\lambda_2S_2+\ldots\;,
\label{eq_seff}
\end{eqnarray}
where the effective couplings $\lambda_i$ are functions of the original parameters, $\lambda_i=\lambda_i(\beta,N_\tau)$.
The introduction of the logarithm is convenient in order to employ the 
graphical cluster expansion, described e.g.~in~\cite{Montvay:1994cy}, which features 
only connected diagrams.

An important observation is that only graphs winding around the time
direction are needed for the effective action.
To see this, consider for example the integration of a cube shown in 
Fig.~\ref{fig_cube}, which is a valid graph with non-vanishing contribution. 
Nevertheless, since it does not wind around the temporal lattice extent, 
the spatial link integrations suffice to remove {\it all} dependence on link 
variables. As a result, the cube contributes only as a function $f(\beta)$ 
to the effective partition function and hence cancels in expectation values or 
renormalised quantities such as the physical free energy density. This is true 
for all graphs which do not wind around the lattice. 

After spatial integration, the interaction terms $S_i$ in 
Eq.~(\ref{eq_seff}) then depend on the link variables only via Polyakov loops
\begin{eqnarray}
 L(\vec{x})\equiv\mathrm{Tr}\;W(\vec{x})\equiv
\mathrm{Tr}\;\prod_{\tau=1}^{N_\tau}U_0(\vec{x},\tau)\;.
\end{eqnarray}
Thus we transform the remaining path integration measure from temporal link 
variables to traced Polyakov loops, which introduces a reduced Haar measure 
denoted by $e^V$. We now consider $SU(3)$ and parametrise the 
measure by two angles of the diagonalised Polyakov loop, providing another factor $e^V$, 
so that at every spatial lattice site $\vec{x}$ we have
\eqa
  && L(\theta,\phi)=e^{i\theta}+e^{i\phi}+e^{-i(\theta+\phi)}\,,\qquad V =\frac{1}{2}\ln \Big(27-18|L|^2+8\Real(L^3)-|L|^4\Big)\nonumber \\
\longrightarrow\quad&&\int\left[\prod_{\tau=1}^{N_\tau}dU_0(\tau)\right]=\int\de W =\int \de L \;e^{V} = \int_{-\pi}^{+\pi}\de\theta\int_{-\pi}^{+\pi}\de\phi \; e^{2V}\;.
\qea

In Eq.~(\ref{eq_seff}), we arrange the interaction terms in ascending
 order of their leading power
of $\beta$ in the strong coupling expansion of the corresponding 
effective coupling.
The first such interaction term is between nearest neighbours $\langle ij\rangle$ in
the fundamental representation and has the form
\begin{eqnarray}
\lambda_1S_1=\lambda_1(\beta,N_\tau)\sum_{<ij>}\Big(L_iL_j^\ast +L_i^\ast L_j\Big)\;,
\label{eq_nnint}
\end{eqnarray}
where $L_i=L(\vec{x}_i)$. This leading order 
contribution comes from $N_\tau$ temporal plaquettes building a chain 
around the lattice, followed by spatial link integration.
\begin{figure}[t]
\centerline{
\includegraphics[width=0.55\textwidth]{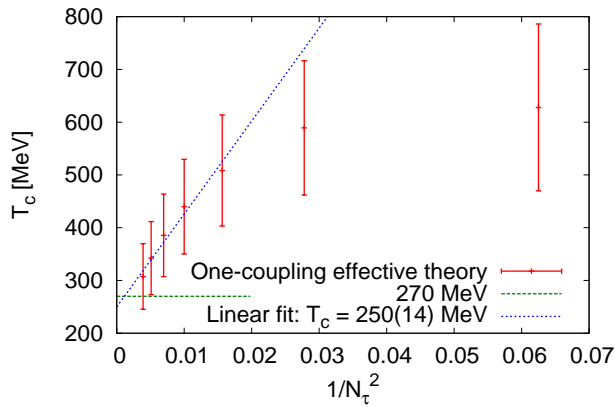}
}
\caption[]{Critical temperature for the pure gauge deconfinement transition, 
extracted from the 3d effective theory, Eqs.~(\ref{efft_YM}, \ref{eq_lambda1}).}
\label{cont}
\end{figure}
Knowledge of the relations 
$\lambda_i(\beta,N_\tau)$ allows to convert the critical
couplings $\lambda_{i,c}$ of the three-dimensional theory to those of 
the original theory. In particular, for the nearest
neighbour coupling $\lambda_1(u,N_\tau)$ we find
\eqa
\hspace{-0.4cm}\lambda_1(u, 2)&=&u^2\exp\bigg[2\left(4u^4+12u^5-18u^6-36u^7\right.\nonumber\\
&&\hspace*{2cm} \left.\left.
+\frac{219}{2}u^8+\frac{1791}{10}u^9+\frac{830517}{5120}u^{10}\right)\right]\;,\nonumber\\
\hspace{-0.4cm}\lambda_1(u, 4)&=&u^4\exp\bigg[4\left(4u^4+12u^5-14u^6-36u^7 \right. \nonumber\\
&&\hspace*{2cm}\left.\left.
+\frac{295}{2}u^8+\frac{1851}{10}u^9+\frac{1035317}{5120}u^{10}\right)\right]\nonumber\;,\\
\lambda_1(u,N_{\tau}\geq5)&=&u^{N_\tau}\exp\bigg[N_{\tau}\bigg(4u^4+12u^5-14u^6-36u^7\nonumber\\
&&\hspace*{2.5cm}
+\frac{295}{2}u^8+\frac{1851}{10}u^9+\frac{1055797}{5120}u^{10}\bigg)\bigg]\;,
	\label{eq_lambda1}
\qea
where the character expansion coefficient $u=u(\beta)= \beta/18+\ldots$ of the 
fundamental representation is used instead of $\beta$ due to better apparent convergence.

Altogether Eq.~(\ref{eq_seff}) has infinitely
many couplings with loops in all irreducible representations, to all powers and at 
all distances. In the strong 
coupling region, the $S_i$ are the more suppressed the higher the index $i$. 
In~\cite{Langelage:2010yr} we saw indeed that the 
influence of the next-to-nearest neighbour coupling as well as the 
adjoint coupling are negligible within the current level of accuracy
when investigating the phase transition. We thus neglect these and higher-order correction terms in the following. Summing up powers of the nearest-neighbour interaction term \cite{Langelage:2010yr}, the effective theory for thermal Yang-Mills reads
\eq
Z_{\mathrm{eff}} = \int \left(\prod_{i} \de L_i  \;e^{V_i}\right)
		\prod_{<ij>}(1+2\lambda_1\Real L_i L^*_j)\;.
\label{efft_YM}
\qe
The deconfinement transition of Yang-Mills theory is reflected in an 
order-disorder transition of
the effective theory.
We determine the critical coupling to be $\lambda_{1,c} = 0.18805(2) \equiv \lambda_0$, cf. Sec.\ref{subsec:flux}\footnote{The slight discrepancy between $\lambda_0$ and the $\lambda_{1,c}$ of \cite{Langelage:2010yr} is due to finite-size effects, as in the latter determination system sizes only up to $N_s=14$ were employed.}, and, by inverting
Eq.~(\ref{eq_lambda1}), extract the $\beta_c(N_\tau)$ which agree with full 4d Monte
Carlo results within better than 10\% accuracy up to $N_\tau=16$. 
Using the non-perturbative beta-function for $a(\beta)$
provided in \cite{Sommer}, these can then be converted to deconfinement temperatures, as shown in 
Fig.~\ref{cont}. Note that all points stem from a single determination of $\lambda_{1,c}$ in the effective
theory. A continuum extrapolation in $a^2$, i.e.~$1/N_\tau^2$, is feasible and predicts a deconfinement transition of 
$T_c=250(14)(50)$ MeV. The second, systematic error is taken as the difference between the $\mathcal{O}(u^{10})$
and $\mathcal{O}(u^{9})$ in Eq.~(\ref{eq_lambda1}).  Encouraged by this result, we now proceed
to extend the effective theory to include heavy Wilson fermions.

\subsection{Heavy fermions: LO hopping parameter expansion}
\label{subsec:hop}

\begin{figure}[t!]
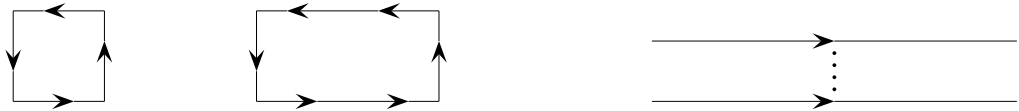

\hspace{1cm}
\scalebox{0.4}{
\psline[arrowsize=0.5 0.5]{->}(0,0)(2,0)
\psline(2,0)(3,0)
\psline[arrowsize=0.5 0.5]{->}(3,0)(3,2)
\psline(3,2)(3,3)
\psline[arrowsize=0.5 0.5]{->}(3,3)(1,3)
\psline(1,3)(0,3)
\psline[arrowsize=0.5 0.5]{->}(0,3)(0,1)
\psline(0,1)(0,0)
\psline[arrowsize=0.5 0.5]{->}(8,0)(10,0)
\psline[arrowsize=0.5 0.5]{->}(10,0)(13,0)
\psline(13,0)(14,0)
\psline[arrowsize=0.5 0.5]{->}(14,0)(14,2)
\psline(14,2)(14,3)
\psline[arrowsize=0.5 0.5]{->}(14,3)(12,3)
\psline[arrowsize=0.5 0.5]{->}(12,3)(9,3)
\psline(9,3)(8,3)
\psline[arrowsize=0.5 0.5]{->}(8,3)(8,1)
\psline(8,1)(8,0)
\psline[arrowsize=0.5 0.5]{->}(21,0)(27,0)
\psline(27,0)(33,0)
\psline[arrowsize=0.5 0.5]{->}(21,2)(27,2)
\psline(27,2)(33,2)
\psdot(27,0.4)
\psdot(27,0.8)
\psdot(27,1.2)
\psdot(27,1.6)
}
\caption{Different terms occurring in the hopping expansion. Left: 
Plaquette. Middle: An example of a 6-link graph. Right: Generalised 
Polyakov loop, i.e. the loop winds around the temporal direction $n$ times 
before the link variables are traced.}
\label{fig_hopex}
\end{figure}

Heavy fermions are conveniently introduced using the hopping parameter 
expansion, 
which is described in \cite{Montvay:1994cy} at zero temperature and 
discussed in \cite{Langelage:2009jb,Langelage:2010yn} for finite 
temperature. The quark part of the action for $N_f$ mass-degenerate 
flavours with masses $M_f=M$ can then be written as
\begin{eqnarray}
-S_q=-N_f\sum_{l=1}^\infty\frac{\kappa ^l}{l}\mathrm{Tr}\,H[U]^l\;,
\qquad\kappa=\frac{1}{2aM+8}\;,
\label{eq_hopex}
\end{eqnarray}
with the hopping matrix
\begin{eqnarray}
H[U]_{y,x}\equiv\sum_{\pm\nu}\delta_{y,x+\hat{\nu}}(1+\gamma_\nu)U_\nu(x)\;,\qquad\gamma_{-\nu}=-\gamma_\nu\;.
\label{eq_hopmat}
\end{eqnarray}
Thus each hop to a neighbouring lattice site gives a power of the hopping 
parameter $\kappa$. The quark chemical potential $\mu$ is introduced as 
usual by a factor $\mathrm{e}^{a\mu}$ ($\mathrm{e}^{-a\mu}$) multiplying link variables in positive (negative) time 
direction \cite{Hasenfratz:1983ba}. The Kronecker delta in 
Eq.~(\ref{eq_hopmat}) requires that the graphs in the hopping expansion be closed. In Fig.~\ref{fig_hopex} we show several graphs appearing in the hopping expansion.
As an example and in order to establish a physical mass as reference scale, we
calculate the pion mass to leading orders,
\eq
	aM_\pi = -2\ln(2\kappa)  - 6\kappa^2 - 54\kappa^4 -24\kappa^2 \frac{u}{1-u} + \mathcal{O}(\kappa^6, \kappa^2u^5)\;.
	\label{eq:pion_mass}
\qe

At finite temperature there are also graphs with a nontrivial winding number, such as 
generalised Polyakov loops
\begin{equation}
 \mathrm{Tr}\,W^n(\vec{x})\equiv\mathrm{Tr}\left(\prod_{\tau=1}^{N_\tau}U_0(\vec{x},\tau)\right)^n\;,\qquad 1\leq n\leq\infty\;,
\end{equation}
winding around the lattice $n$ times before being traced, cf. 
Fig.~\ref{fig_hopex} (right).

We obtain the effective action from the full partition function in 
the same way as in pure gauge theory,
\begin{eqnarray}
Z&=&\int[dU_0][dU_i]\exp\left[-S_g-S_q\right]
=\int[dU_0]\exp\left[-S_{\mathrm{eff}}\right]\;,\nonumber\\
-S_{\mathrm{eff}}&=&\ln\int[dU_i]\exp\left[-S_g-S_q\right]\;.
\end{eqnarray}
We are now faced with a double series expansion in $u(\beta)$ and 
$\kappa$, i.e.~the effective couplings depend on both parameters and $N_\tau$. Furthermore, quarks of finite mass lead to terms in the action 
which explicitly break the $Z(3)$ symmetry present in the pure gauge case. 
The effective action may then be written as
\begin{eqnarray}
-S_{\mathrm{eff}}=\sum_{i=1}^\infty\lambda_i(u,\kappa,N_\tau)S_i^s-
2N_f\sum_{i=1}^\infty\left[h_i(u,\kappa,\mu,N_\tau)S_i^a+
\bar{h}_i(u,\kappa,\mu,N_\tau)S_i^{a,\dagger}\right]\;.
\label{eq_defseff}
\end{eqnarray}
The $\lambda_i$ are defined as the effective couplings of the 
$Z(3)$-symmetric terms $S_i^s$, whereas the $h_i$ multiply the asymmetric 
terms $S_i^a$.
Consequently, only the latter are $\mu$-dependent and we recover pure 
gauge theory for $\kappa\rightarrow0$, as in the full theory. We have not 
included the factor $2N_f$ in the definition of the $h_i$, since 
there are $N_f$ mass-degenerate quarks with $2$ spin degrees of freedom, 
all giving the same contribution. The $h_i$ and $\bar{h}_i$ are related via
\begin{eqnarray}
 \bar{h}_i(u,\kappa,\mu, N_\tau)=h_i(u,\kappa,-\mu, N_\tau)\;.
\end{eqnarray}

We shall now derive combined strong coupling and 
hopping parameter expansions of these effective couplings, again by 
employing the graphical cluster expansion.
Similar to the case of pure gauge theory, graphs contributing to the cluster expansion 
have to wind around the lattice in the compact $\tau$-direction. Hence, the leading order contributions are 
Polyakov loops, and we can read off the leading-order couplings $h_1$ and $\hb_1$:
\begin{eqnarray}
h_1S_1^a+\bar{h}_1S_1^{a,\dagger}=-\sum_{\vec{x}}\Big[(2\kappa e^{a\mu})^{N_\tau}L(\vec{x}
)+(2\kappa e^{-a\mu})^{N_\tau}L^\ast(\vec{x}
)\Big] \; \rightarrow \;
h_1 = (2\kappa e^{a\mu})^{N_\tau}.
\label{eq:loh}
\end{eqnarray}
Note the minus sign which is due to the anti-periodic boundary conditions for fermions.
It is possible to sum up the contributions of all generalised 
Polyakov loops  oriented in positive time direction,
\begin{eqnarray}
\exp\left\lbrace\!\!-2N_f\sum_{\vec{x}}\!\sum_{n=1}^\infty\Bigg[\frac{(-1)^n}{n}(2\kappa e^{a\mu})^{nN_\tau}\mathrm{Tr}\;
\left(W^n\right)
\Bigg]\!\!\right\rbrace\!=\!\prod_{\vec{x}}\!\det\!\Big[1+(2\kappa e^{a\mu})^{N_\tau}W\Big]^{2N_f},
\end{eqnarray}
using $\exp\mathrm{Tr}\ln A\equiv\det A$,  and similarly for 
the conjugate loop. The  
three-dimensional effective action to leading order in the hopping expansion then corresponds to the static approximation and reads
\begin{eqnarray}
Z_{\mathrm{eff}} =
\int [\de U_0]
\Bigg(\!\prod_{<ij>}\Big[ 1 + 2\lambda_1\Real L_i^* L_j \Big]\Bigg)\!\!\Bigg(\!\prod_{\vec{x}}
\det \Big[ \Big(1+h_1W_{\vec{x}}\Big)\!\left(1+\bar{h}_1 W_{\vec{x}}^\dagger\right) \Big]^{2N_f}\!\Bigg).
\label{efft+kappa}
\end{eqnarray}

\subsection{Heavy fermions: beyond leading order}

\label{subsec:hop2}
\begin{figure}[t]
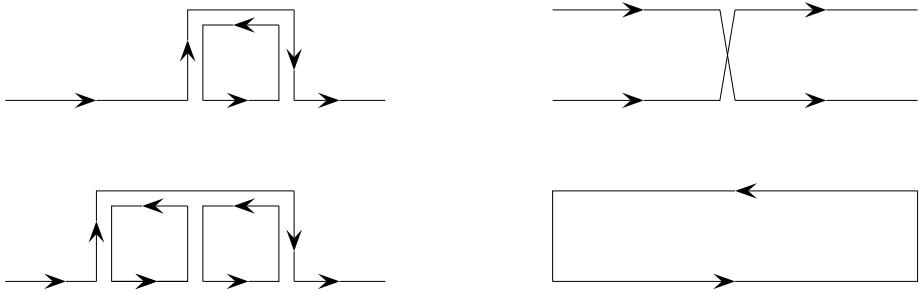

\vspace{3cm}
\hspace{1.5cm}
\scalebox{0.4}{
\psline[arrowsize=0.5 0.5]{->}(0,6)(3,6)
\psline(3,6)(6,6)
\psline[arrowsize=0.5 0.5]{->}(6,6)(6,8)
\psline(6,8)(6,9)(9.5,9)
\psline[arrowsize=0.5 0.5]{->}(9.5,9)(9.5,7)
\psline(9.5,7)(9.5,6)
\psline[arrowsize=0.5 0.5]{->}(9.5,6)(11,6)
\psline(11,6)(12.5,6)
\psline(7.5,8.5)(6.5,8.5)(6.5,6)
\psline[arrowsize=0.5 0.5]{->}(6.5,6)(8,6)
\psline(8,6)(9,6)(9,8.5)
\psline[arrowsize=0.5 0.5]{->}(9,8.5)(7.5,8.5)
\psline[arrowsize=0.5 0.5]{->}(18,6)(21,6)
\psline(21,6)(23.5,6)(24,9)
\psline[arrowsize=0.5 0.5]{->}(24,9)(27,9)
\psline(27,9)(30,9)
\psline[arrowsize=0.5 0.5]{->}(18,9)(21,9)
\psline(21,9)(23.5,9)(24,6)
\psline[arrowsize=0.5 0.5]{->}(24,6)(27,6)
\psline(27,6)(30,6)
\psline[arrowsize=0.5 0.5]{->}(0,0)(2,0)
\psline(2,0)(3,0)
\psline[arrowsize=0.5 0.5]{->}(3,0)(3,2)
\psline(3,2)(3,3)(9.5,3)
\psline[arrowsize=0.5 0.5]{->}(9.5,3)(9.5,1)
\psline(9.5,1)(9.5,0)
\psline[arrowsize=0.5 0.5]{->}(9.5,0)(11,0)
\psline(11,0)(12.5,0)
\psline(7.5,2.5)(6.5,2.5)(6.5,0)
\psline[arrowsize=0.5 0.5]{->}(6.5,0)(8,0)
\psline(8,0)(9,0)(9,2.5)
\psline[arrowsize=0.5 0.5]{->}(9,2.5)(7.5,2.5)
\psline[arrowsize=0.5 0.5]{->}(3.5,0)(5,0)
\psline(5,0)(6,0)(6,2.5)
\psline[arrowsize=0.5 0.5]{->}(6,2.5)(4.5,2.5)
\psline(4.5,2.5)(3.5,2.5)(3.5,0)
\psline[arrowsize=0.5 0.5]{->}(3.5,0)(5,0)
\psline(5,0)(6,0)
\psline[arrowsize=0.5 0.5]{->}(18,0)(24,0)
\psline(24,0)(30,0)(30,3)
\psline[arrowsize=0.5 0.5]{->}(30,3)(24,3)
\psline(24,3)(18,3)(18,0)
}
\vspace{0.25cm}
\caption{Left: Graphs reducing to Polyakov loops after spatial link integration resulting in ${\cal{O}}(\kappa^2u^l)$ corrections, with 
$1\leq l\leq N_\tau-1$. Right: Graphs 
of ${\cal{O}}(\kappa^{2N_\tau+2})$, leading to interactions between 
Polyakov loops at distance $a$ after spatial link integration.}
\label{fig_kappacorr}
\end{figure}

Corrections to the static approximation come from graphs that also contain spatial link 
variables. To order ${\cal{O}}(\kappa^4)$ there is the plaquette term, see Fig.~\ref{fig_hopex} (left). Its contribution can be absorbed in the gauge term leading to a shift 
in $\beta$ and hence to a $\kappa$-dependence of $u$,
\begin{eqnarray}
\beta \rightarrow \beta+48N_f\kappa^4\quad \Rightarrow\quad u(\beta)\rightarrow u(\beta,\kappa)\;.\label{eq:ushift}
\end{eqnarray}
Higher-order graphs containing 6 and more 
links~\cite{Montvay:1994cy}, for which Fig.~\ref{fig_hopex} (middle) is an 
example, may be neglected to the orders to which we have calculated our effective 
couplings.

Next, we consider leading-order corrections to the winding graphs in the full (3+1)-dimensional action 
and observe their effect on the 3d effective theory, 
cf.~Fig.~\ref{fig_kappacorr}. Graphs on the left lead to 
${\cal{O}}(\kappa^2)$ corrections to the couplings $h_1, \bar h_1$ after spatial link integration. 
We have calculated the first correction to be
\begin{eqnarray}
(2\kappa e^{a\mu})^{N_\tau}6N_\tau \kappa^2\sum_{l=1}^{N_\tau-1}u^l= 
(2\kappa e^{a\mu})^{N_\tau}6N_\tau \kappa^2\frac{u-u^{N_\tau}}{1-u}\,,
\end{eqnarray}
by summing over $1\leq l\leq N_\tau-1$ possible attached plaquettes.
As a result we get the higher order version of the effective coupling,
\begin{eqnarray}
h_1=(2\kappa e^{a\mu})^{N_\tau}\left(1+6N_\tau \kappa^2 \frac{u-u^{N_\tau}}{1-u}+\ldots\right)\;.
\label{eq:h1_renorm}
\end{eqnarray}
Including all corrections up to ${\cal{O}}(u^n\kappa^m)$, with $n+m=7$, we obtain
\begin{eqnarray}
 h_1(u,\kappa, N_\tau\geq 3)=(2\kappa e^{a\mu})^{N_\tau}\;&\exp&\left[6N_\tau\kappa^2u \left(\frac{1-u^{N_\tau-1}}{1-u}\right.\right.\nonumber\\
 &+&\left.4u^4-8\kappa^2+8\kappa^2u+4\kappa^2u^2-4\kappa^4\bigg)\right]\;,
\label{eq_kappafinal}
\end{eqnarray}
where we used the fact that terms $\sim N_\tau^2$ or higher can be resummed by writing 
the correction as an exponential. The graph in the lower right of Fig.~\ref{fig_kappacorr}  
contributes a $\kappa$-dependent term to the already included nearest-neighbour interaction of 
Eq.~(\ref{eq_nnint}). This exemplifies how $\lambda_1(u,N_\tau)$ becomes 
$\lambda_1(u,\kappa,N_\tau)$. Note that these corrections do not change the form of 
the effective action Eq.~(\ref{efft+kappa}).

By contrast, a graph that gives rise to a new term in the effective 
action is shown in the upper right corner of Fig.~\ref{fig_kappacorr}. Together with its 
oppositely oriented counterpart it implies interactions between
Polyakov loops of the same orientation,
\begin{eqnarray}
&&h_2S^a_2+\bar h_2S^{a,\dagger}_2 = \sum_{<ij>}\Big(h_2L_iL_j+\bar{h}_2L_i^\ast L_j^\ast\Big)\;,\label{h2}\\
&&h_2=(2\kappa e^{a\mu})^{2N_\tau}\frac{N_\tau\kappa^2}{N_c}\;,
\qquad\bar{h}_2=(2\kappa e^{-a\mu})^{2N_\tau}\frac{N_\tau\kappa^2}{N_c}\;.\nonumber
\end{eqnarray}
However, the terms arising from both graphs on the right of Fig.~\ref{fig_kappacorr} are parametrically of high order 
${\cal{O}}\left(\kappa^{2N_\tau+2}\right)$ and will thus be neglected along with
even higher-order corrections.

For the remainder of this work we thus approximate thermal lattice 
QCD with heavy quarks by the effective theory Eq.~(\ref{efft+kappa}) with the couplings 
$h_1$ of Eq.~(\ref{eq_kappafinal}) and $\lambda_1$ of Eq.~(\ref{eq_lambda1}) with $u(\beta,\kappa)$
as in Eq.~(\ref{eq:ushift}). Since we retain only one coupling of each sort, let us drop the index 1
and use the index $i$ only to label lattice sites to lighten the notation. Finally,  
we rewrite $Z_\mathrm{eff}$ in a form more convenient for its numerical
evaluation. A useful identity for the determinants is
\eqa
	\det(1+h W) &=& 1+h\tr W + h^2 \tr W^\dagger + h^3 = 
		1 + h L + h^2 L^* + h^3\;,\nonumber \\
	\det(1+\hb W^\dagger) &=& 1+\hb\tr W^\dagger + \hb^2 \tr W + \hb^3 = 
		1 + \hb L^* + \hb^2 L + \hb^3\;.\nonumber
\qea
The heavy quark contribution per flavour and site $\vec x_i$ then becomes
\eq
Q_i(h,\bar{h}) =\left[(1 + h L_i + h^2 L_i^* + h^3)(1 + \hb L_i^* + \hb^2 L_i + \hb^3)\right]^2\,,\label{eq:Qdef}
\qe
and the 3d effective theory for thermal QCD with heavy quarks reads simply
\eq
Z_{\mathrm{eff}} = \int\!\left(\prod_{i} \de L_i \;e^{V_i}Q^{N_f}_i(h,\bar{h}) \right)
		\prod_{<ij>}(1+2\lambda \Real L_i L^*_j)\;.
\label{efft_w_Q}
\qe

In our numerical investigations we consider the partition function Eq.~(\ref{efft_w_Q}) for \mbox{$N_f=1$}.
Since the deconfinement transition at high temperature happens at small $h$, we can recover an arbitrary 
number of flavours by using the approximation
\eq
	\det (1 + h W)^{N_f} \approx \exp{(N_f h L)}\;,\;h(N_f) \approx \frac{1}{N_f}h(N_f=1)\;\;.
\qe

\subsection{Observables}

In our numerical simulations we are interested in the phase structure.
With a standard action linear in its couplings, one would typically construct observables using 
energy- and magnetisation-type fields defined as
\eq
	E_{lin} = \frac{1}{3N_s^3}\sum_{<ij>}2 \Real L_iL_j^*\,,\quad
	Q_{lin} = \frac{1}{N_s^3} \Big| \sum_i L_i \Big| \label{eq:linobs}\;
\qe
on finite volumes $V = N_s^3$.
In our case the non-linear action implied by Eq.~(\ref{efft_w_Q}) suggests to use a different definition, closer to what actually drives the dynamics of the system,
\eq
	E = \frac{1}{\lambda} \frac{1}{3N_s^3}\sum_{<ij>} \log \Big( 1 + 2\lambda \Real L_iL_j^* \Big)\,,\quad
	Q = \frac{1}{h}\frac{1}{N_s^3}\sum_i \log |Q_i|\;.
	\label{eq:basic_observables_definition}
\qe
These are proportional to 
the previous definitions in the limit of small couplings $(\lambda, h,\bar h)$.
For the finite size scaling analyses of the phase structure we used the susceptibility and the Binder cumulant
constructed from the observables $O \in \{E, Q, E_{lin}, Q_{lin}\}$,
\eq
	\chi_O = N_s^3(\avg{O^2}-\avg{O}^2)\,,\quad B_{4,O} = \frac{\avg{(O-\avg{O})^4}}{\avg{(O-\avg{O})^2}^2}\;.
	\label{eq:susc_deltabinder_definition}
\qe

\subsection{Flux representation and numerical simulation}
\label{subsec:flux}

The system described by the partition function Eqs.~(\ref{efft+kappa}, \ref{efft_w_Q}) has  a complex action and thus suffers from a sign problem for $h\neq\bar{h}$ ($\mu > 0$). In the simplified case where the local variables $L_i$ take only values in the center,  $L_i \in Z(3)$, i.e.~the Potts model in a magnetic field, the sign problem can be solved by using a cluster algorithm~\cite{Alford:2001ug} or a change of variables 
to obtain a flux representation \cite{Condella:1999bk, Alford:2001ug}. 
The latter approach can be generalised to the case of 
$SU(3)$-valued Polyakov loops as was done in~\cite{Gattringer:2011gq} for a related model. 
The flux representation of the partition function reads 
\eq
	Z_\mathrm{eff}(\lambda, h,\bar h) = \sum_{\{\substack{n_b,m_b\\ n_i,m_i}\}} \prod_{b=(i,\mu)} W_b(n_b,m_b)
		\prod_i W_i(p_i,q_i,n_i,m_i)\,.\label{eq:Z_worm}
\qe
As the $SU(3)$-valued variables have been integrated out, the degrees of freedom are now local incoming and outgoing 
currents, $n_{i,\mu},m_{i,\mu} = 0,1$, respectively, located on the links connecting the site $i$ and $i+\mu$, $\mu=\pm1,\ldots,\pm d$ 
as well as charges (or monomers) $n_i, m_i = 0,\ldots,4$ located on the sites of our cubic lattice. 
$W_b(n_b,m_b)$ is given by
\eq
W_b(n_b,m_b) = 
\left\{
\begin{array}{ll}
1,& \text{ if } (n_b, m_b) = (0,0)\\
\lambda,& \text{ if } (n_b, m_b) = (0,1) \text{ or } (1,0)\\
0& \text{otherwise}\,.
\end{array}
\right.
\qe
If we rewrite the factor $Q$ in Eq.~(\ref{eq:Qdef}) as a power series in $L,L^\ast$, $Q = \sum_{n,m} \xi_{n,m} L^nL^{\ast m}$ then the site weight $W_i$ reads
\eq
	W_i(p_i,q_i,n_i,m_i) = \xi_{n_i,m_i}\int\!\mathrm{d}U\;L^{q_i+n_i}L^{\ast p_i+m_i}\geq 0\;,
\qe
where the last expression contains a traced $SU(3)$ link integral given in closed form in~\cite{Eriksson:1980rq,Carlsson:2008dh}.
$W_i$ is positive for all $\mu\geq 0$ and the model no longer has a minus sign problem. Locally, this integration creates a Potts constraint
$q_i+n_i =   p_i+m_i \text{ mod } 3$, where
\eq
q_i = \sum_{\mu = \pm 1,\ldots,\pm d}n_{i,\mu}\,,\quad p_i = \sum_{\mu =\pm 1,\ldots,\pm d}m_{i,\mu}\,.
\qe
Thus, local currents are conserved modulo 3.

The partition function Eq.~(\ref{eq:Z_worm}) is well-suited for an application of the worm 
algorithm~\cite{Prokof'ev:2001zz} and its variants. To enable sampling of configurations 
in the presence of an external magnetic field ($h,\bar{h}\neq 0$) we implemented 
a variant of the algorithm presented in~\cite{Gabriel2002}.
Having changed the degrees of freedom, the observables Eqs.~(\ref{eq:basic_observables_definition}) have to be re-expressed. 
The Polyakov loop and its complex conjugate are now given by
\eq
	\frac{1}{V} \left\langle\sum_i L_i\right\rangle =  \frac{1}{V}\frac{\partial}{\partial h}\ln Z \approx \langle n/h\rangle, \quad
	 \frac{1}{V} \left\langle\sum_i L^\dagger_i\right\rangle \approx\langle m/\bar h\rangle\,,
\qe
 where $\langle n\rangle$ and $\langle m\rangle$ denote the average number of monomers of either type. The relations become
exact only in the limit $h,\bar h\rightarrow 0$, due to the non-exponential form of the quark part of expression 
Eq.~(\ref{efft+kappa}). The quark density is given by 
\eq
n_q = \frac{1}{V N_\tau}\frac{\partial}{\partial \mu} \ln Z = \frac{1}{VN_\tau}\left\langle\sum_i \frac{\partial_\mu W_i}{W_i}\right\rangle .
\qe
\begin{figure}[t]
\centerline{
\includegraphics[width=0.5\textwidth]{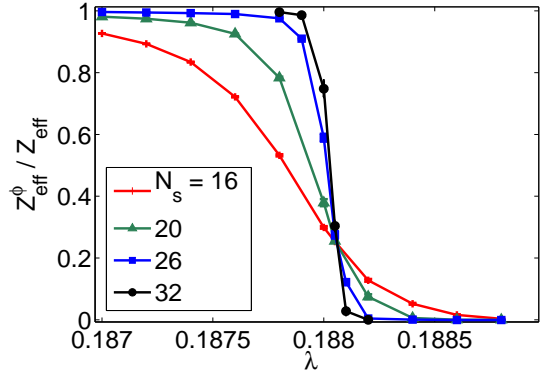}
}
\caption[]{$Z^{\phi}_{\mathrm{eff}}/Z_{\mathrm{eff}}$ as a function of $\lambda$ for the pure gauge case, crossing at $\lambda_0 \approx 0.18805$.}
\label{fig:worm_foo}
\end{figure}

The worm algorithm relies on the sampling of the 2-pt function 
$G(i,j) = \langle L_iL_j^\ast\rangle$ rather than the partition function Eq.~(\ref{eq:Z_worm}). 
$G(i,j)$ can be estimated 
during a worm update~\cite{Prokof'ev:2001zz} as $G(i,j) = h(i,j) / Z$, where $h(i,j)$ corresponds to the histogram of configurations with
the worm head (say $L^\ast$) at site $j$ and its tail ($L$) at site $i$.

An observable more suitable for the flux representation is the free energy of an interface enforced in the broken phase by twisted boundary conditions in the e.g. $z$-direction, $L_{i+N_s\mathbf{e}_z} = \mathrm{e}^{i\phi} L_i$ with $\phi = {\frac{2\pi}{3} q}, q = 0,1,2$. 
The partition function with twisted boundary conditions is 
\eq
\displaystyle
Z^{\phi}_{\mathrm{eff}}= \sum_{\{\substack{n_b,m_b\\n_x,m_x}\}} \mathrm{e}^{i\phi\sum_{x\in P} \left[n_z(x)-m_z(x)\right]}\prod_b W_b\prod_i W_i\;,
\qe
where $j_z = \sum_{x\in P} \big(n_z(x)-m_z(x)\big)$ is the flux through the plane $P = \{\vec x~|~z = N_s-1\}$.
If we consider the case $h=\bar h = 0$, i.e.~the model representing pure gauge
theory, then the spontaneous breaking of centre symmetry is signalled by the ratio
\eq
\frac{Z^{\phi}_{\mathrm{eff}}}{Z_\mathrm{eff}} = 
\left\{
\begin{array}{ll}
1,& \text{ for } T< T_c \\
0,& \text{ for } T > T_c\;.
\end{array}
\right.
\qe
In Fig.~\ref{fig:worm_foo} we show $Z^{\phi}_{\mathrm{eff}}/Z_{\mathrm{eff}}$ as a function of $\lambda$ for $\phi=2\pi/3$. 
Since our model has a weak first order transition several volumes cross at the transition point
$\lambda_0 = 0.18805(2)$.

On the other hand, compared to the full theory the sign problem
in the standard representation of the effective theory, Eq.~(\ref{efft_w_Q}), is mild, 
similar to the case of the Potts model in an 
external field~\cite{Kim:2005}.
\begin{figure}[t]
\centerline{
\includegraphics[width=0.5\textwidth]{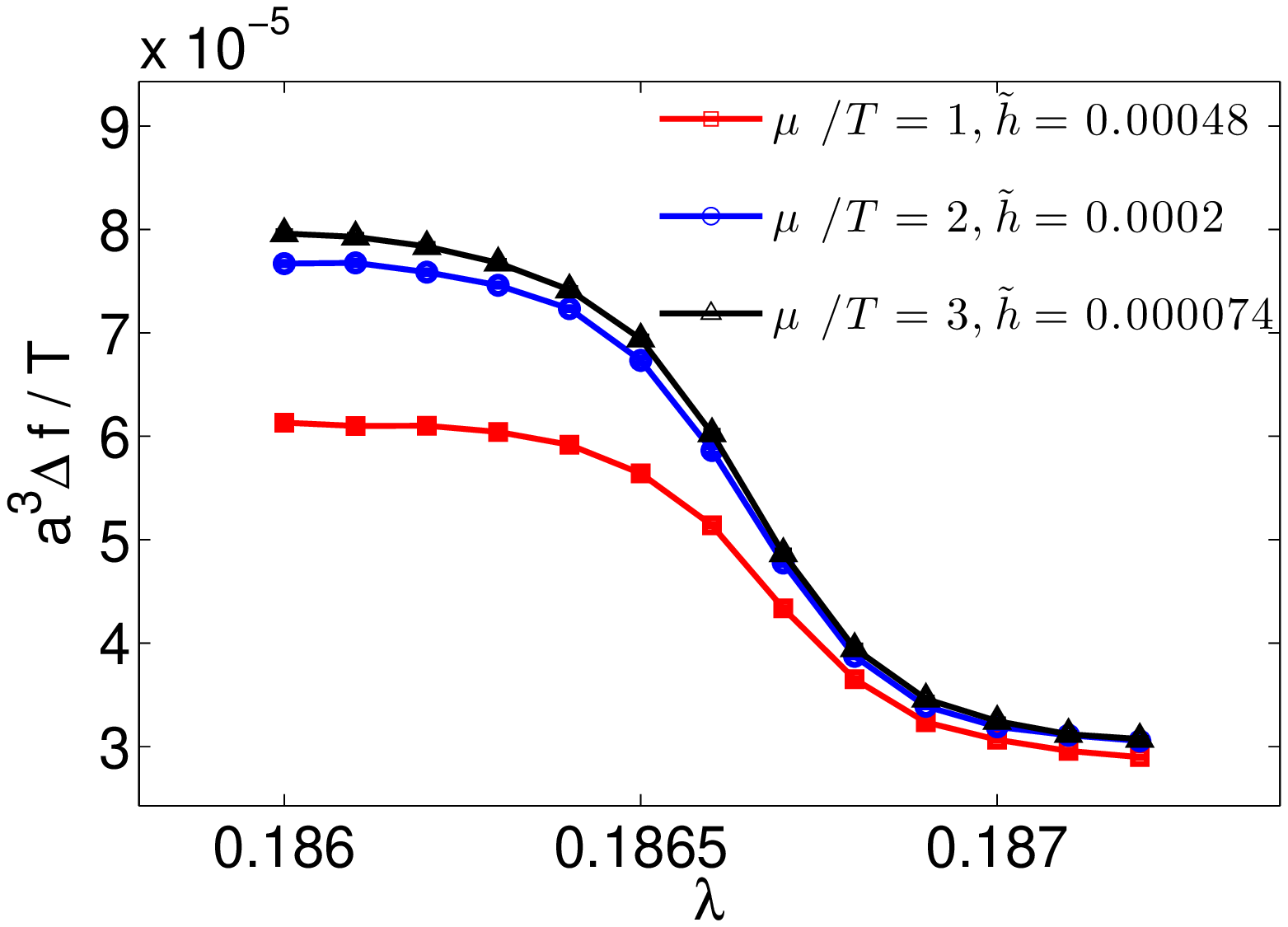}
\includegraphics[width=0.5\textwidth]{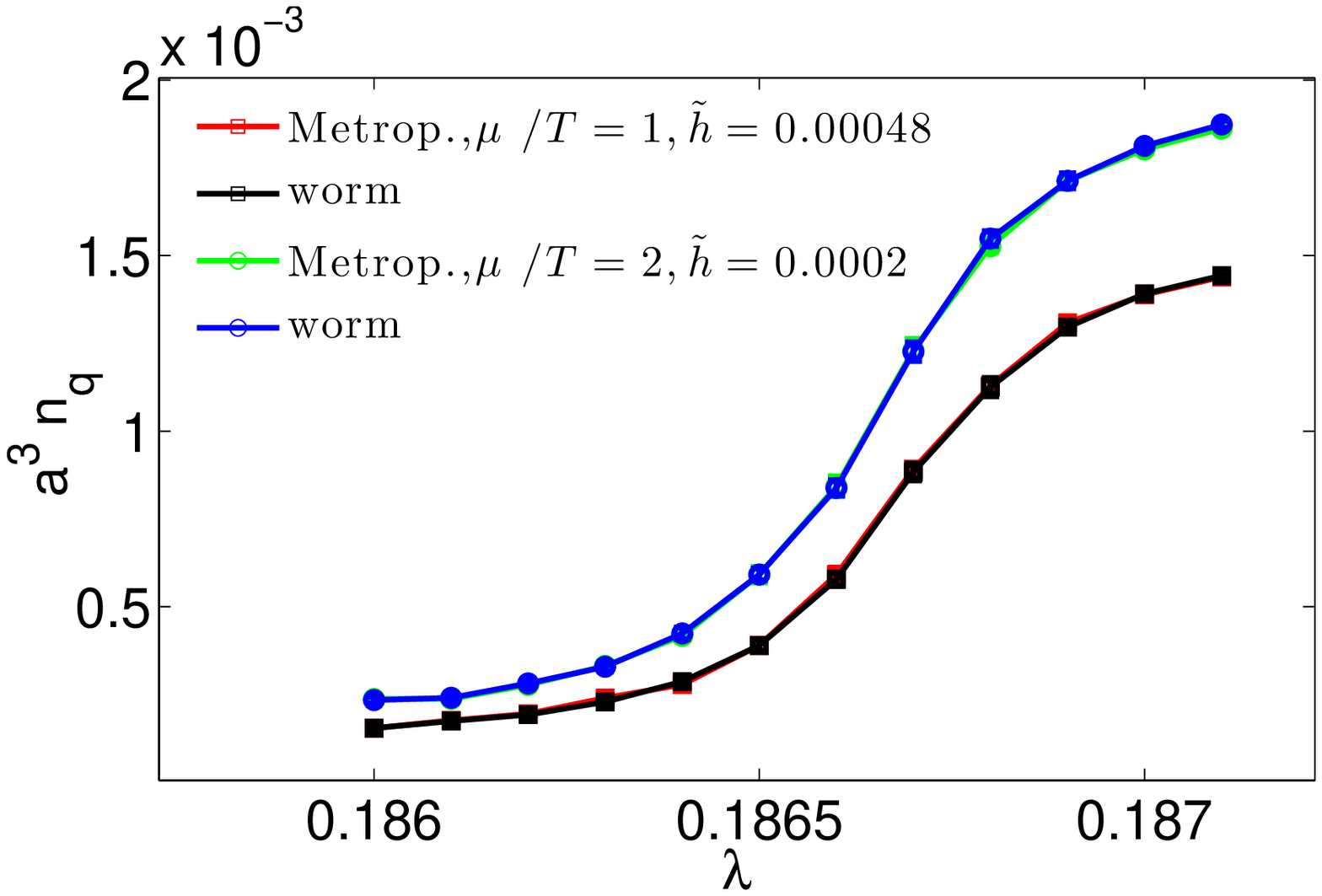}
}
\caption[]{Left: Average sign, Eq.~(\ref{eq:avsign}), of the fermion determinant in the effective theory in the vicinity of the critical point for various chemical
potentials on a $24^3$ lattice. Right: Quark density calculated with $Z_\mathrm{eff}$ from Eq.~(\ref{efft_w_Q}) (Metropolis) or Eq.~(\ref{eq:Z_worm}) (worm) on $24^3$ lattices.}
\label{fig:algs}
\end{figure}
Using the modulus of the determinant in a Metropolis algorithm while 
reweighting in its phase, system sizes up to $N_s=24$ for values of the chemical potential up to $\mu/T \sim 3$ can be reached, 
with a fully controlled average sign. To demonstrate this, we monitor the ratio of the full and phase 
quenched partition functions,
\eq
\langle \mbox{sign}\rangle_{||}=\frac{Z_\mathrm{eff}}{Z_{\mathrm{eff}||}}=e^{-\frac{V}{T}\Delta f(\mu^2)}\;, \quad Z_{\mathrm{eff}||}: \mbox{phase quenched}.\label{eq:avsign}
\qe 
The corresponding difference in free energy density is a volume-independent
measure for the strength of the sign problem and is shown in 
Fig.~\ref{fig:algs} (left). Even for the largest system sizes to be used in our scaling analyses,
the average sign remains significant and fully controlled up to large chemical potentials $\mu/T\sim 3$. 
This is corroborated by comparing a physical observable such as the quark density between
the worm algorithm (without sign problem) and standard Metropolis algorithm with Polyakov loops and
reweighting in the phase of the determinants. 
Fig.~\ref{fig:algs} (right) shows that no difference is discernible between
the two ways of evaluating the observable.

We thus conclude that the sign problem can be fully controlled and even solved for our effective theory. Since the observables of interest are more readily accessible in the original degrees of freedom, we have mainly used the Metropolis algorithm for the following numerical simulations.

\section{The deconfinement transition for heavy quarks}
\label{sec:res}
\subsection{Zero baryon density}

\begin{figure}[t]
\centerline{
\includegraphics[width=0.4\textwidth]{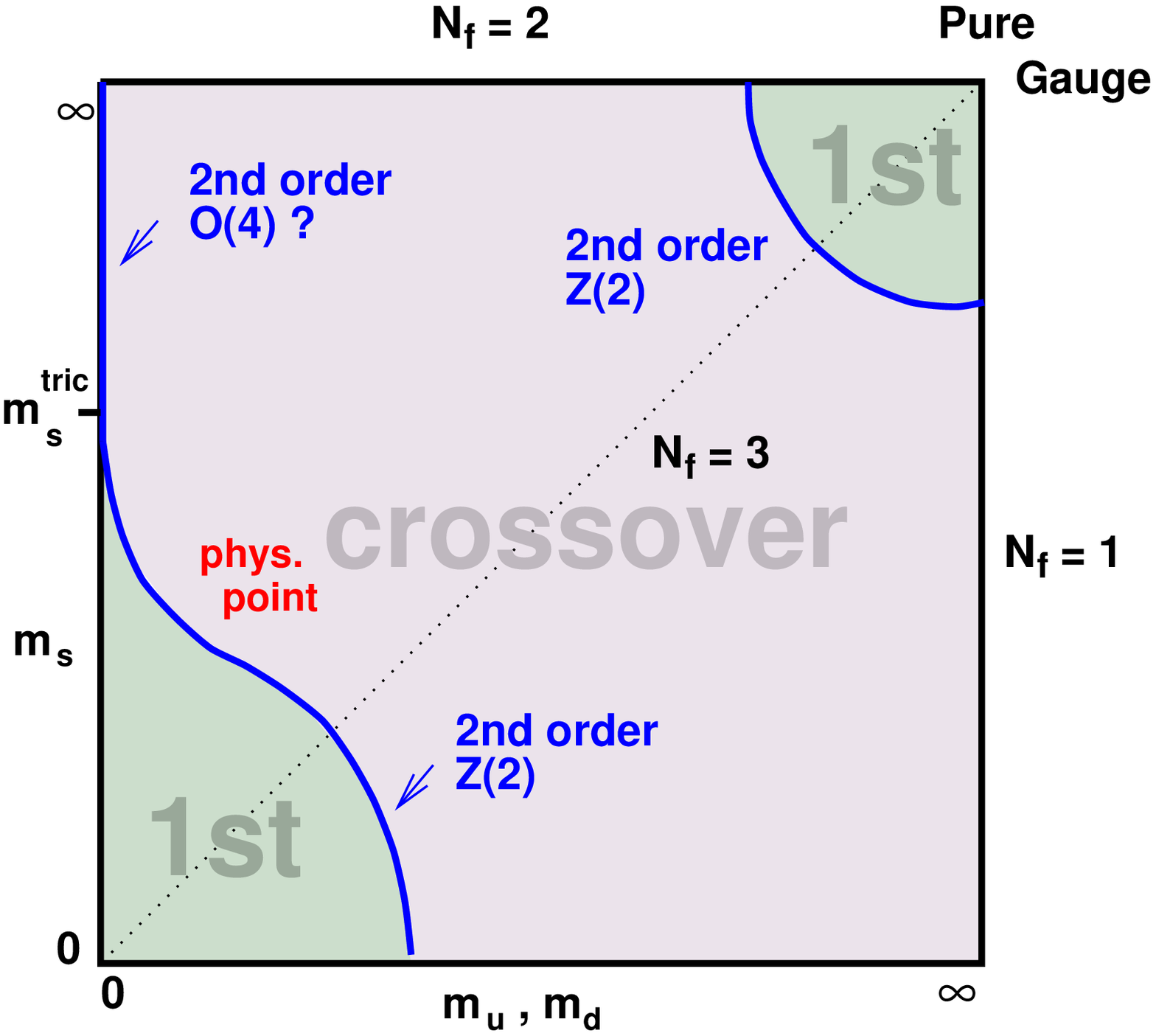}\quad
\includegraphics[width=0.4\textwidth]{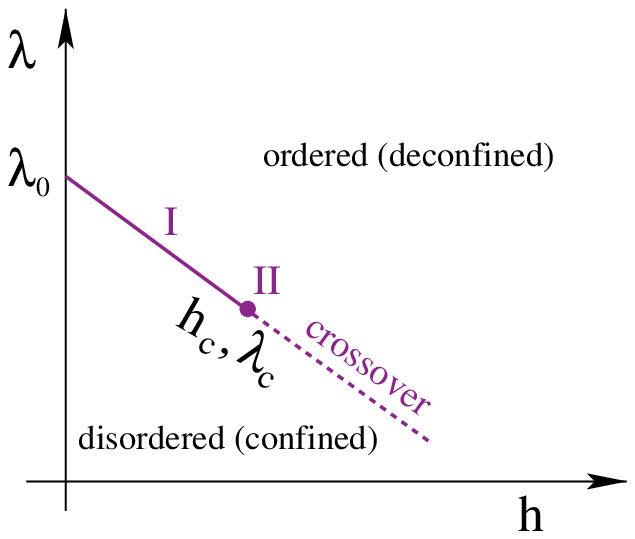}
}
\caption[]{Schematic phase diagram for $\mu=0$. Left: Nature of the QCD thermal transition for different quark masses. Right: Expected phase diagram for the effective theory corresponding to $\mu=0$.}
\label{schem}
\end{figure}

As a first application of the effective theory we investigate the deconfinement transition 
of QCD with heavy quarks as a function of quark mass and chemical potential. 
We begin by considering the case of zero baryon density,
shown schematically in Fig.~\ref{schem} (left). In the pure gauge limit, the deconfinement transition is 
of first order. Dynamical quarks at 
any fixed $N_f$ break the global $Z(3)$
symmetry of the QCD action explicitly.  As a consequence, the phase transition weakens with decreasing quark 
masses until it vanishes at a critical point. For still lighter quarks the deconfinement transition is an
analytic crossover. 

This behaviour is inherited by the effective theory. For a given $N_f$ and $\mu=0$, 
we have $h=\bar{h}$ and the effective theory has two couplings, $(\lambda, h)$.
The first-order phase transition of the one-coupling theory extends to a first-order 
line with a weakening transition as $h$ increases. Eventually the transition should vanish 
at a critical point, as sketched in Fig.~\ref{schem} (right). These expectations are based
on the known results of the 3d 3-state Potts model in an external field \cite{Alford:2001ug, Kim:2005}, which shows the same symmetry breaking pattern.
While the behaviour of the system in the vicinity of the critical point is dictated by the universality class, the location
of the transition in parameter space, in particular the critical parameters where it changes its nature, are not. Hence, our investigation will give valuable additional
information on QCD. Previous investigations to locate the critical heavy lattice quark mass 
have been made on coarse $N_\tau=4$ lattices for $N_f=1$~\cite{Alexandrou:1998wv}, and 
in~\cite{Saito:2011fs} for several flavours.
           
\begin{figure}[t]
\centerline{
\includegraphics[height=0.4\textwidth]{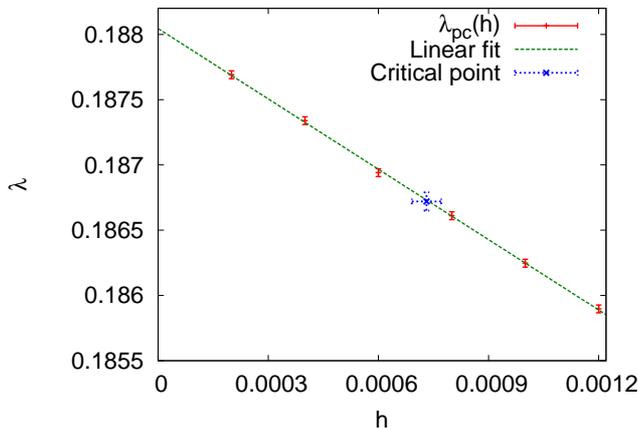} 
}
\caption{The pseudo-critical line or phase boundary in the thermodynamic limit. Also the critical point, Eq.~(\ref{eq:muzero_critpoint}), is shown.}
\label{fig:lpc}
\end{figure}

In order to determine the phase diagram Fig.~\ref{schem} (right), we follow a two-step procedure.
First we determine the phase boundary, i.e.~the pseudo-critical line $\lambda_{pc}(h)$ in the two-coupling space of the effective theory. In a second step, using dedicated finite size scaling analyses,
we determine the order of the transition along that line, and in particular the location $(\lambda_c,h_c)$ 
of the critical point.
In order to accomplish the first task we fix the external field variable to the values 
$h = 0.0002 - 0.0012$ on lattice sizes
$N_s = 10-24$, and then scan for the corresponding pseudo-critical coupling $\lambda_{pc}$.  
As indicators for the phase boundary we use maxima of susceptibilities and minima of Binder cumulants constructed from the observables given in Eqs.~(\ref{eq:linobs}, \ref{eq:basic_observables_definition}).
We extrapolate these to infinite volume using
\eq
	\lambda_{pc}(h,N_s) = \lambda_{pc}(h)+c_1(h)/N_s^\alpha\;.
\qe
\begin{figure}[t]
\centerline{
\includegraphics[height=0.4\textwidth]{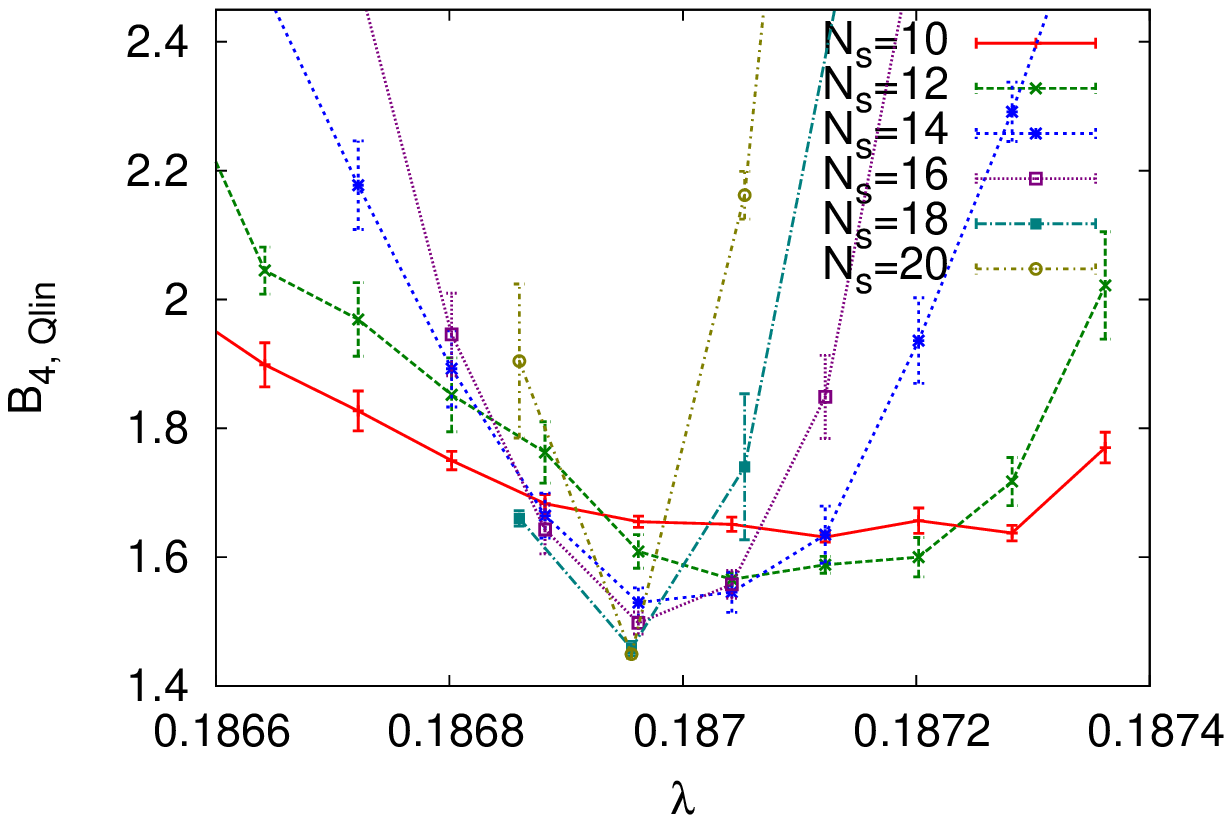}
\includegraphics[height=0.4\textwidth]{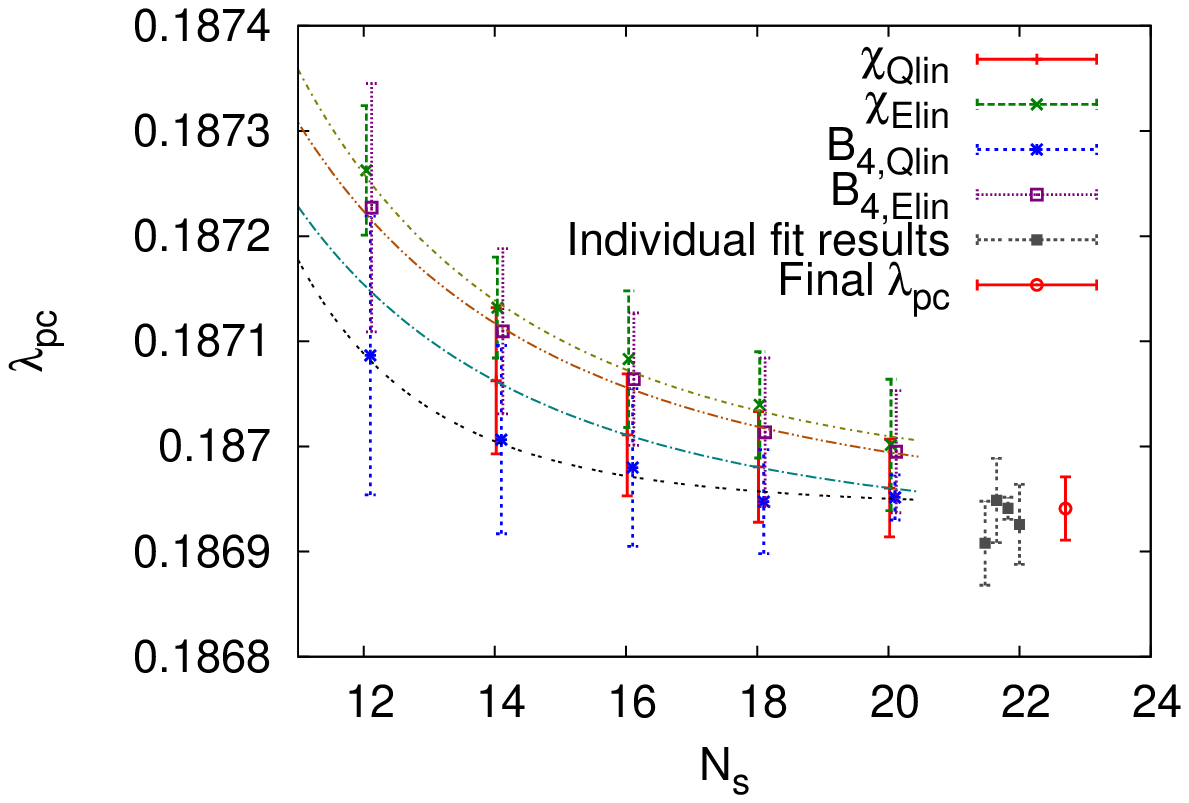}
}
\caption{Pseudo-critical couplings for $h = \bar h = 0.0006$. Left: $B_{4,Q_{lin}}$ on various volumes, 
the minima approach $\lambda_{pc}(h)$ in the thermodynamic limit. Right: 
Extrapolation of $\lambda_{pc}(h,N_s)$ from several observables to infinite volume.
}
\label{fig:detcritline}
\end{figure}
For each coupling and system size we generated at least $10^5$ 
configurations.
Fig.~\ref{fig:detcritline} shows the minima of the Binder cumulant 
$B_{4,Q_{lin}}$ for $h=0.0006$ and the extrapolation of these values 
along with those of other observables.
This results in the pseudo-critical line shown in Fig.~\ref{fig:lpc}. It is well described by a linear 
fit, due to the small magnitude of $h$ and the argument given in~\cite{Alford:2001ug}: 
along the line of first  order transitions the free energy densities of 
the disordered (confined) phase and ordered (deconfined) phase 
are equal, $f_c(\lambda,h)=f_d(\lambda,h)$.
Expanding both sides about the pure gauge transition, $(\lambda_0,h=0)$, 
and noting that $\partial_h f_c = \langle L\rangle = 0$ in zero external field,  we obtain
\eq
	\lambda_{pc}(h) = \lambda_0 - a_1 h,\quad a_1 = \frac{\partial_h f_d}{\partial_\lambda( f_c-f_d)}\bigg|_{(\lambda_0,0)}\,. \label{eq:linear_fit}
\qe
A fit with $\chi^2=0.26$ yields $a_1=1.797(18)$ and $\lambda_0 = 0.18805(1)$ in very good agreement with the estimate from the flux
representation,
see Fig.~\ref{fig:worm_foo} and Sec.~\ref{subsec:flux}. An alternative approach to pin down the pseudo-critical line $\lambda_{pc}(h)$ 
would be to determine the principal axes $\mathcal{E},\mathcal{M}$ of the joint probability distribution $P(E, Q)$ of 
our variables.  The critical line is then defined by a 
certain symmetry condition on $P(\mathcal{M})$, demanding the vanishing of the third moment 
$\langle \mathcal{M}^3\rangle=0$. We have checked explicitly around the critical point that the two 
approaches give consistent results.

\begin{figure}[h]
\centerline{
\includegraphics*[height=0.4\textwidth]{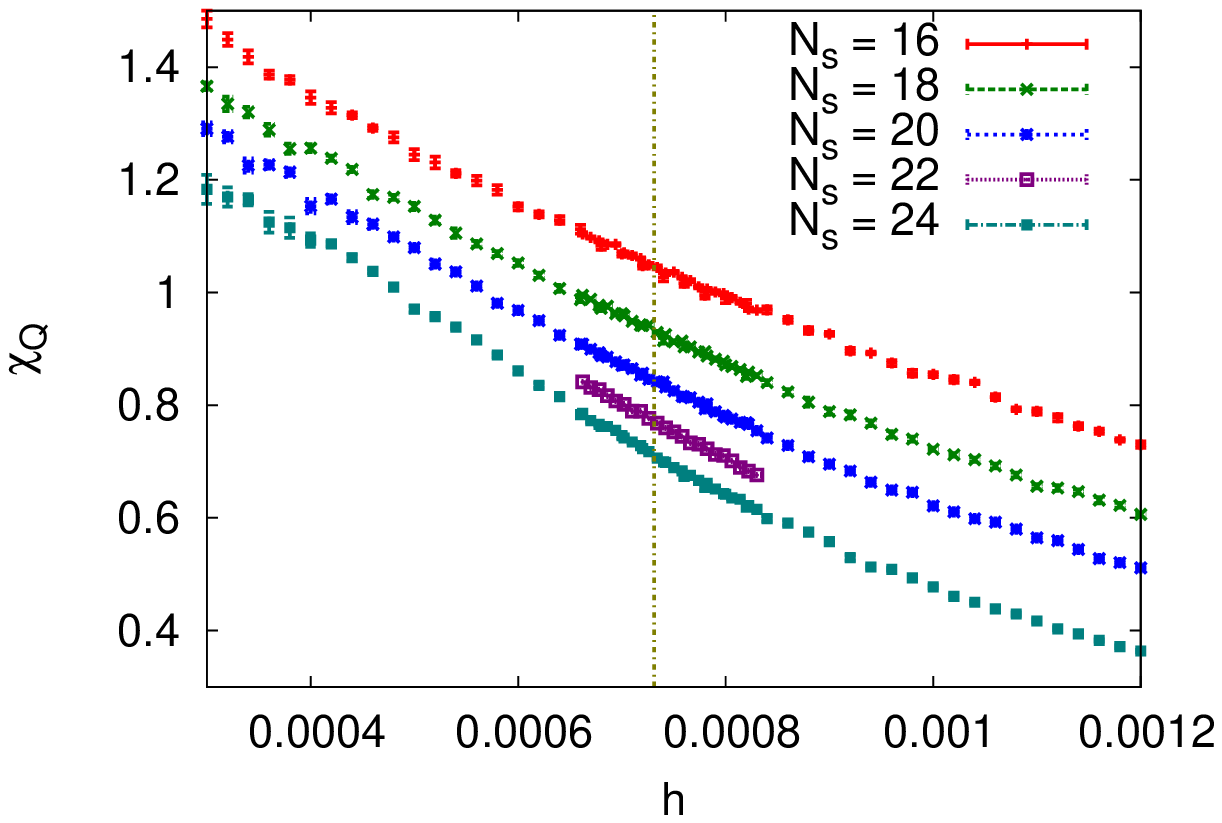}
\includegraphics*[height=0.4\textwidth]{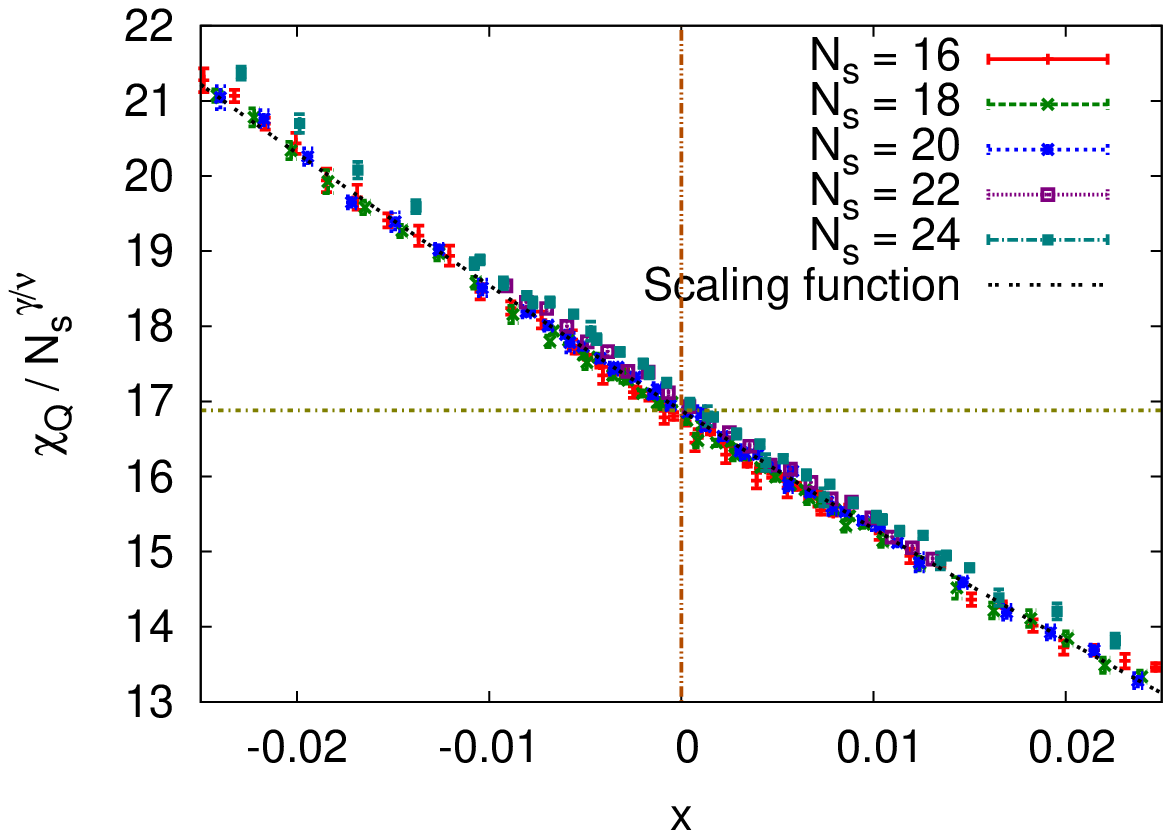} 
}
\caption{Left: $\chi_Q$ as a function of $h_{pc}(\lambda)$ for several volumes. Right: Same, now with rescaled 
axes, Eq.~(\ref{eq:fss}), to produce a collapsing curve described by the universal scaling function (see Table \ref{tab:mu0_fits}). 
The vertical line marks the critical $h$, the horizontal line is $f_0$.}
\label{fig:chi_Q_mu0}
\end{figure}

\begin{figure}[h]
\centerline{
\includegraphics*[height=0.4\textwidth]{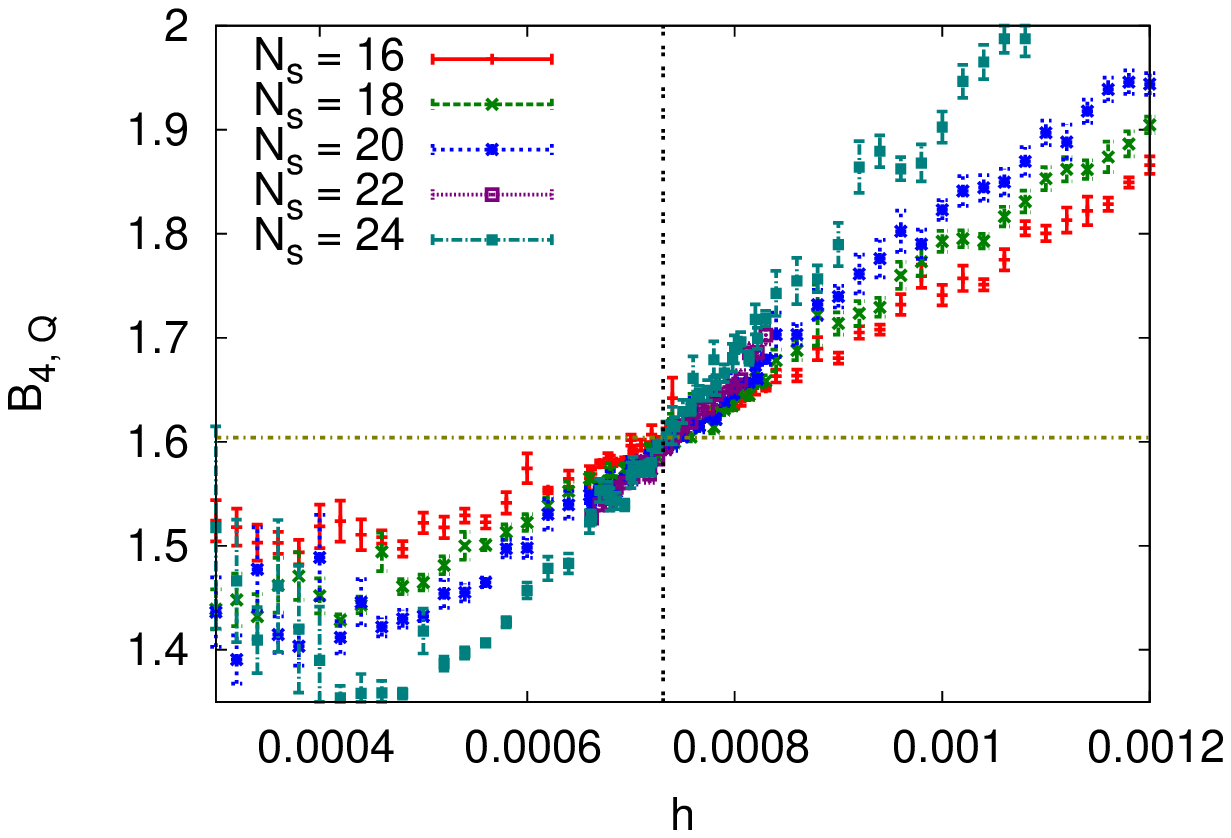}
\includegraphics*[height=0.4\textwidth]{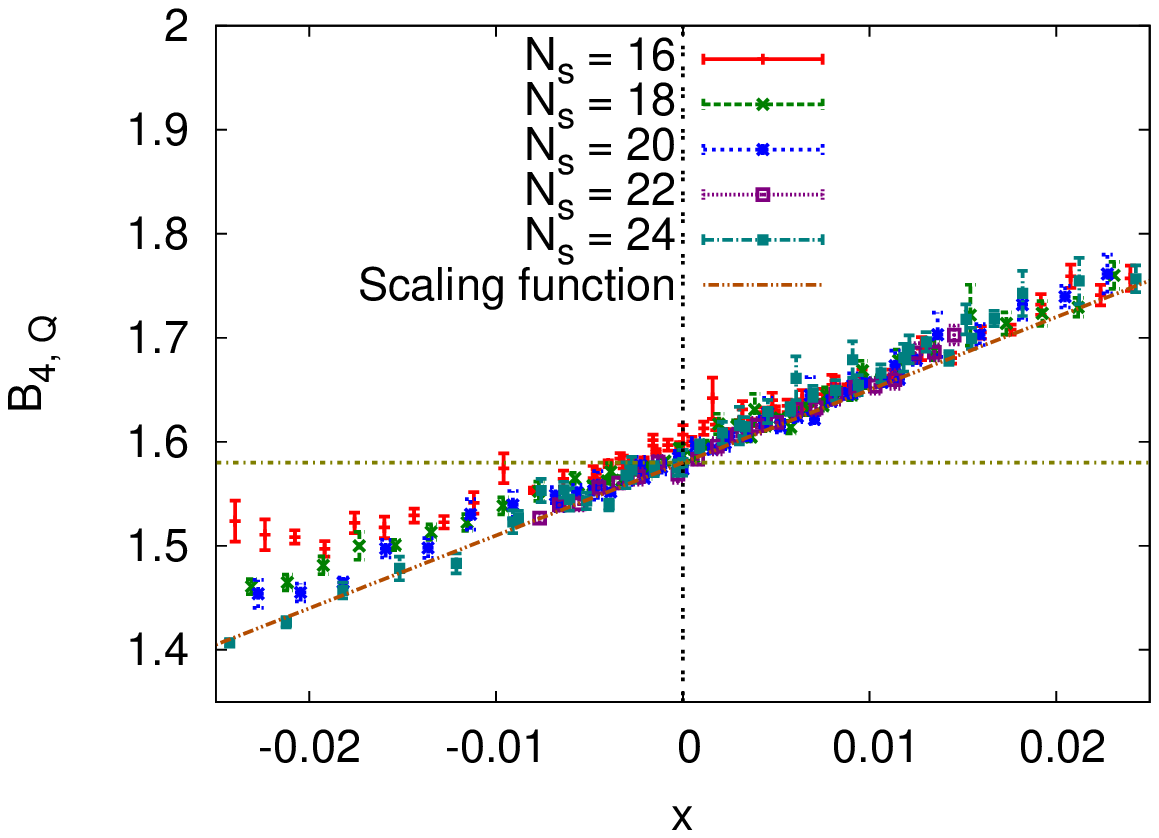} 
}
\caption{Left: $B_{4,Q}$ as a function of $h_{pc}(\lambda)$ for several volumes. The dashed horizontal line denotes the 3d Ising 
value $1.604$ which is the approximate crossing value of the largest volumes $N_s = 20,22,24$, the vertical line marks the critical point. Right: Same, now with rescaled horizontal axis. 
The horizontal line marks $B_{4,Q}(0)$ as found numerically. 
Also shown is the universal scaling function obtained by the fit listed in Table \ref{tab:mu0_fits}, last column.}
\label{fig:dB_4_Q_mu0}
\end{figure}

In order to locate the critical end point of the first-order line and to establish its universality class, 
we study finite size scaling of the data taken along the pseudo-critical line. 
Close to criticality our observables should scale according to 
\eq
	\chi_Q = N_s^{\gamma/\nu}f_{\chi_Q}(x)\label{eq:fss}\,,\quad B_{4,Q}=f_{B_{4,Q}}(x)\,,\quad x\equiv(h-h_c)N_s^{1/\nu}
\qe
with dimensionless scaling functions $f_O$,
provided we move along the tangent $h_{pc}(\lambda)$.
In the vicinity of the critical point the scaling functions may thus be expanded,
\eq
f_O(x) = f_0+f_1x+f_2x^2+\ldots\,,\label{eq:fitfunc}
\qe
which is the form to which we fit our data.
\begin{table}[t]
\begin{center}
\begin{tabular}{|c||r|r||r|r|}
\hline
	 & $\chi_Q$ & $\chi_Q$ & $B_{4,Q}$ & $B_{4,Q}$  \\
\hline
\hline
	$h_c$			&   $0.00073(1)$		& $0.000739(1)$ 	&	$0.00071(1)$	&	$0.00072(2)$ 	\\
	$\nu$			&   $0.63(1)$			& $0.630$ (fixed) 	&	$0.64(2)$		&	$0.630$ (fixed) 	\\
	$\gamma/\nu$		&	$2.00(1)$			& $1.998(1)$		&	--			&	-- 				\\
	$f_0$			&   $16.88(4)$  			& $16.87(2)$		&	$1.58(1)$		&	$1.58(2)$		\\
	$f_1$			&   $-162(12)$  			& $-157(3)$		&	$7.6(8)$		&	$7.0(2)$		\\
	$f_2$			&   $460(60)$   			& $0$ (fixed)		&	$1(1)	$	&	$0$ (fixed)		\\
\hline
\end{tabular}
\caption{Critical end point $h_c$ from fits to the scaling forms Eqs.~(\ref{eq:fss}) using data
from $N_s = 20,22,24$ lattices (all with acceptable $\chi^2/{\rm dof} < 1.5$). The known 3d Ising critical exponents are $\gamma/\nu=1.962(3)$ and $\nu = 0.6302(1)$~\cite{deng_blote_2003}.}
\label{tab:mu0_fits}
\end{center}
\end{table}
We simulated lattice sizes $N_s = 20,22,24$ with statistics of $\sim7\cdot 10^5$ configurations per parameter set and used binning analyses to control autocorrelations.
For the susceptibility $\chi_Q$, we find $\gamma/\nu$ consistent with the expected 3d Ising value, see 
Table \ref{tab:mu0_fits} and Fig.~\ref{fig:chi_Q_mu0}. 
The fit was repeated fixing $f_2=0$, $\nu=0.630$ (3d Ising), and
varying the fit range $|x|<0.01,0.02,0.03$, with an overall stable outcome. 
For the Binder cumulant $B_{4,Q}$, $f_0$ should approach a 
value characterising its universality class, ($f_0=1.604$ for 3d Ising \cite{deng_blote_2003}).
The same fitting procedure as above was then applied to the $B_{4,Q}$ data for $N_s\geq 20$, with compatible values for $h_c$ and $f_0,\nu$, cf.~Table \ref{tab:mu0_fits}. 
The collapse of the data onto a universal curve under the appropriate rescaling is also shown in 
Figs.~\ref{fig:chi_Q_mu0}, \ref{fig:dB_4_Q_mu0}.


\begin{table}[]
\begin{center}
\begin{tabular}{|c|c|c||c|c|}
\hline
	 $N_f$ & $M_c/T$ & $\kappa_c(N_\tau=4)$ & $\kappa_c(4)$, Ref.~\cite{Saito:2011fs} & $\kappa_c(4)$, Ref.~\cite{Alexandrou:1998wv} \\
\hline
	1 & 7.22(5)  & 0.0822(11)           & 0.0783(4) & $\sim$ 0.08 \\
	2 & 7.91(5)  & 0.0691(\phantom{0}9) & 0.0658(3) & -- \\
	3 & 8.32(5)  & 0.0625(\phantom{0}9) & 0.0595(3) & --  \\
\hline
\end{tabular}
\caption{Location of the critical point for $\mu=0$ and $N_\tau=4$. The first two columns report our results 
(we used for consistency the leading-order relation Eq.~\ref{eq:loh}),
the last two
compare with existing literature.}
\label{tab:m_over_t_nf_comparison}
\end{center}
\end{table}
We are then ready to identify the critical point in Fig.~\ref{fig:lpc} (right) 
\eq
	\Big(\lambda_c = 0.18672(7), h_c = 0.000731(40)\Big)\;.
	\label{eq:muzero_critpoint}
\qe
The values of $\lambda_c, h_c$ can be converted into those of the couplings $\beta_c,\kappa_c$ using
Eqs.~(\ref{eq_lambda1}, \ref{eq_kappafinal}). In order to compare with previous work, we approximate $M_c/T$ with the relation, valid for heavy fermions to leading order in the hopping expansion \cite{green_karsch_84},
\eq
	\exp\Big( - \frac{M}{T}\Big) \simeq \frac{h}{N_f}\;.
	\label{eq:kappa_vs_m-over-t}
\qe
The results are collected in Table \ref{tab:m_over_t_nf_comparison} and are in reasonable agreement with the corresponding ones 
from simulations of 4d QCD with Wilson fermions~\cite{Saito:2011fs, Alexandrou:1998wv} at $N_\tau=4$.

\begin{figure}[t]
\centerline{
\includegraphics*[height=0.4\textwidth]{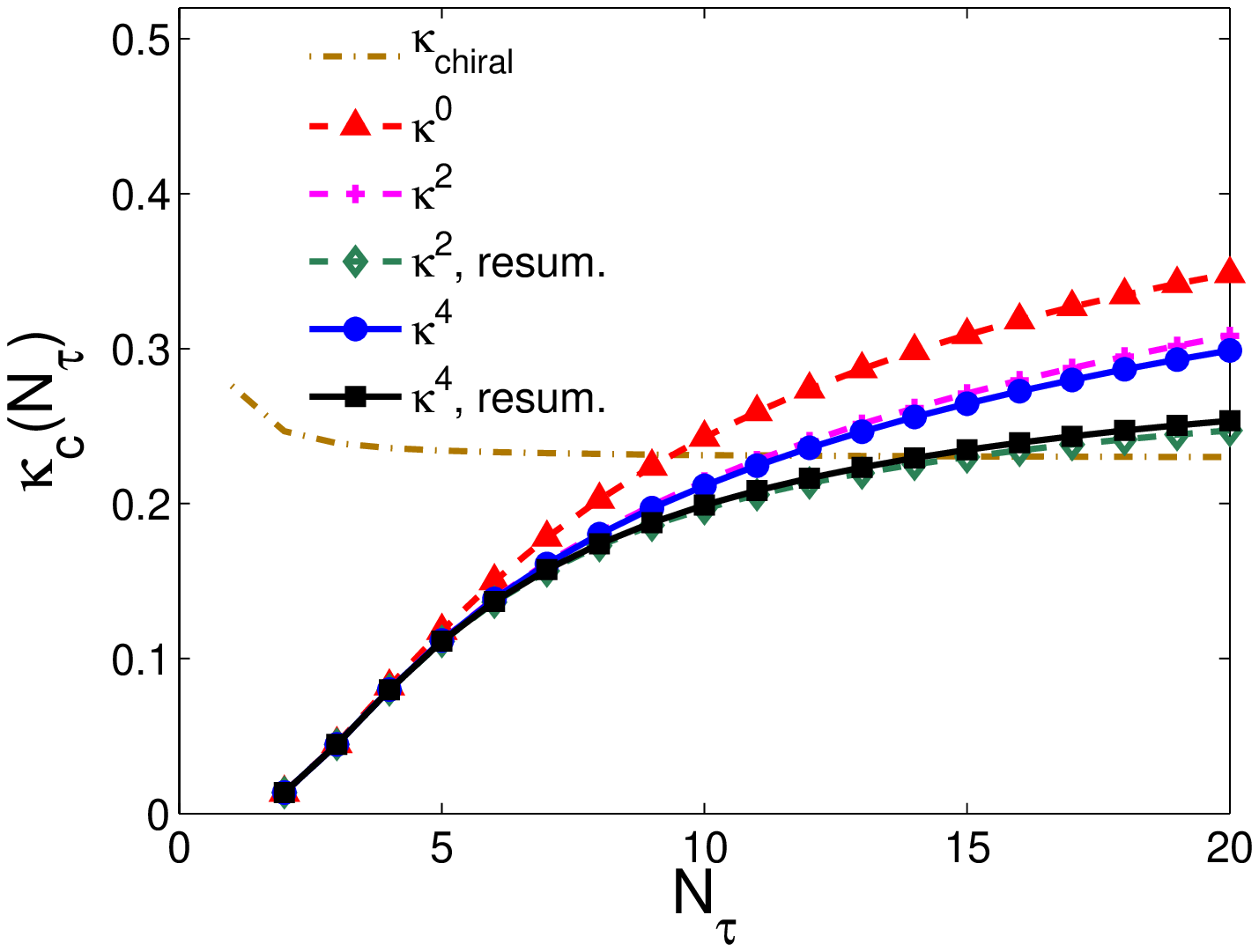}
\includegraphics*[height=0.4\textwidth]{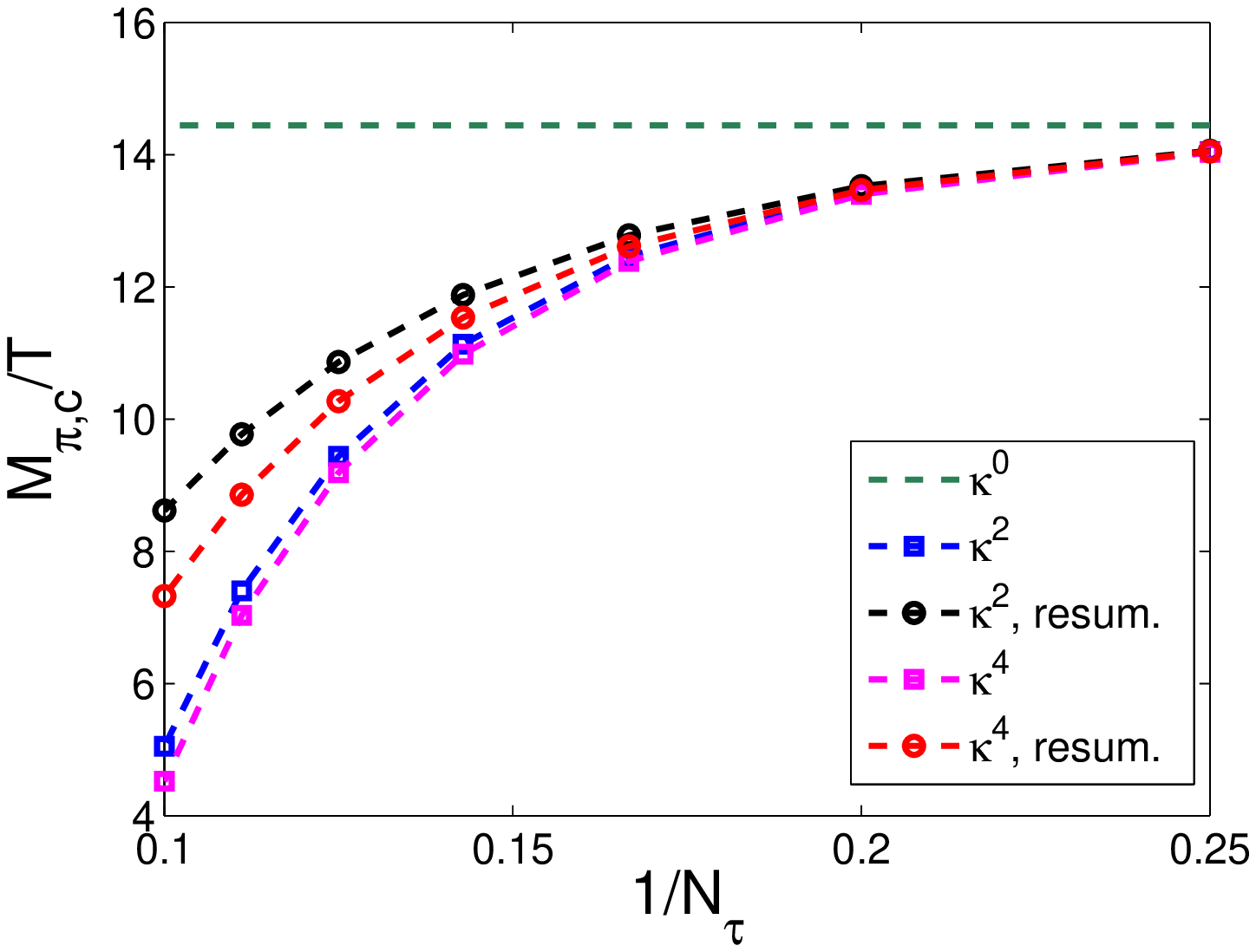}
}
\caption[]{Critical hopping parameter $\kappa_c(N_\tau)$ (left) and critical pion mass $M_{\pi,c}(\kappa_c,u)/T$ (right) given by Eq.~(\ref{eq:pion_mass}) for the end point of the 
$N_f=1$ effective theory to different orders of the hopping expansion.}
\label{fig:kappac}
\end{figure}

As in the case of pure gauge theory, our mapping of the critical effective
couplings to those of QCD can in principle be done for any $N_\tau$, thus
providing predictions for larger $N_\tau$ which have not yet been
simulated in 4d. However, before doing so we need to check how far we can
trust our hopping parameter expansion. Fig.~\ref{fig:kappac} (left) shows the
predictions of the effective theory for $\kappa_c(N_\tau)$ to the orders
$\kappa^2$, $\kappa^4$, resummed and unresummed.
Also shown is the chiral critical hopping parameter, defined by the vanishing of the pion mass Eq.~(\ref{eq:pion_mass}),
and evaluated for the critical gauge couplings,
$\kappa_{ch}\left(u_c(N_\tau)\right)$. Since we are
expanding around infinite quark masses, self-consistency requires
$\kappa_c\ll \kappa_{ch}$. 
Whereas the leading order soon crosses the
$\kappa_{ch}$ line, the corrections are significantly below the
leading order and the exponentiated versions further improve on this. 
Furthermore, literature tells us that $\kappa_{ch}(\beta=0)=0.25$
\cite{Kawamoto:1981} and $\kappa_{ch}(\beta\rightarrow\infty)=0.125$
\cite{Montvay:1994cy}, when all orders are taken into account. Hence, we conclude
that $N_\tau=6$ is the finest thermal
lattice for which our $\kappa_c$ is still significantly smaller than $\kappa_{ch}$ evaluated at
the same gauge coupling, and has not yet crossed the continuum-extrapolated $\kappa_{ch}$.
This is
corroborated by the pion mass $M_{\pi}(u_c,\kappa_c)$ evaluated at
the critical point, shown in Fig.~\ref{fig:kappac} (right).
We observe that beyond
$N_\tau\sim6$ the differences between the non-trivial orders $\kappa^2$
and $\kappa^4$ grow larger, indicating that we leave the convergence
region.

These findings are to be contrasted with $\beta_c(N_\tau)$ of the pure
gauge effective theory, which are within a 10$\%$ range from the known 4d
results up to $N_\tau=16$. However, this is quite natural, since we have
only three non-trivial orders in the hopping expansion, which are
additionally truncated at a low order in $u$, compared to five orders in
the strong coupling expansion. While we hope to extend our results to
larger $N_\tau$ by going to higher orders in $\kappa$, at the moment we
cannot take a continuum limit in the fermionic sector but consider our
results valid up to $N_\tau\sim 6$.

Within this range of validity, we may now discuss the sensitivity of the deconfinement critical line in
Fig.~\ref{schem} (left)
on the cut-off.  The critical pion mass in units of temperature marking the
boundary of the first order deconfinement region shrinks slightly from $N_\tau = 4$ to $N_\tau = 6$. This effect is even smaller in absolute units ($T_c$ also decreases with increasing $N_\tau$), 
in contrast to the critical pion mass evaluated on the chiral critical line, which shrinks by almost a
factor of two \cite{fpk}. Higher orders in the hopping expansion are needed for a definite statement in our case.

\subsection{Finite baryon density}

We now study the deconfinement transition at finite baryon density. For $\mu\neq 0$, we
have $h\neq\bar{h}$ and need to consider the full
parameter space of the effective theory, $(\lambda,h,\bar{h})$. The diagram in 
Fig.~\ref{schem} (right) turns three-dimensional, with a surface of first order phase transitions
terminating in a critical line. Since we are interested in the change of the critical quark mass with chemical potential,   
we prefer to map out the critical line by fixing different chemical potentials and then 
scan for the critical $\kappa$. It is thus convenient to introduce the parameter
\eq
	\tilde{h}\equiv he^{-\mu/T} \left( = (2\kappa)^{N_\tau} \text{ to leading order in $\kappa$} \right)
\qe
and to present our data in the parameter space $(\lambda,\hti,\mu/T)$.

\begin{figure}
\centerline{
\includegraphics*[height=0.4\textwidth]{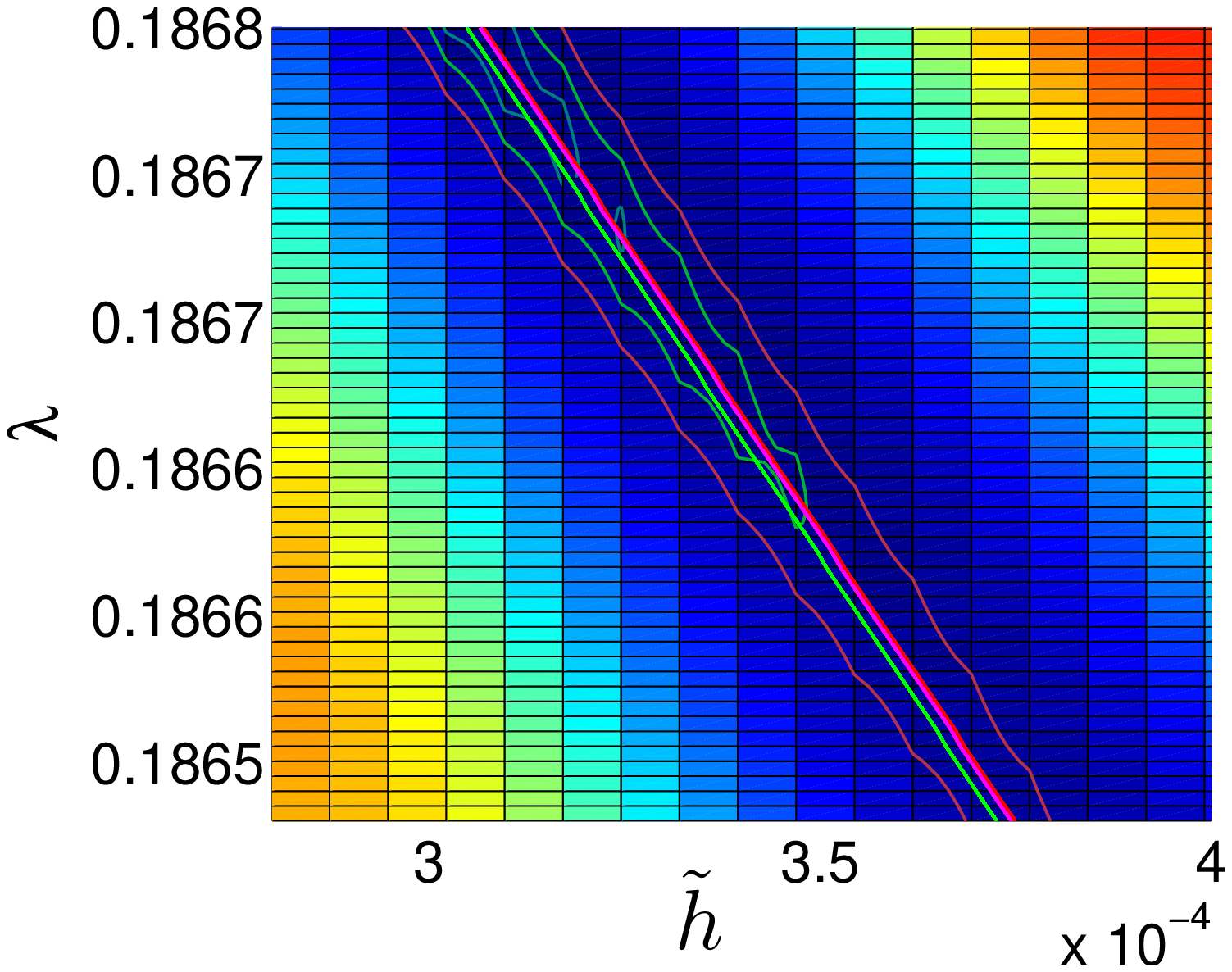}
\includegraphics*[height=0.4\textwidth]{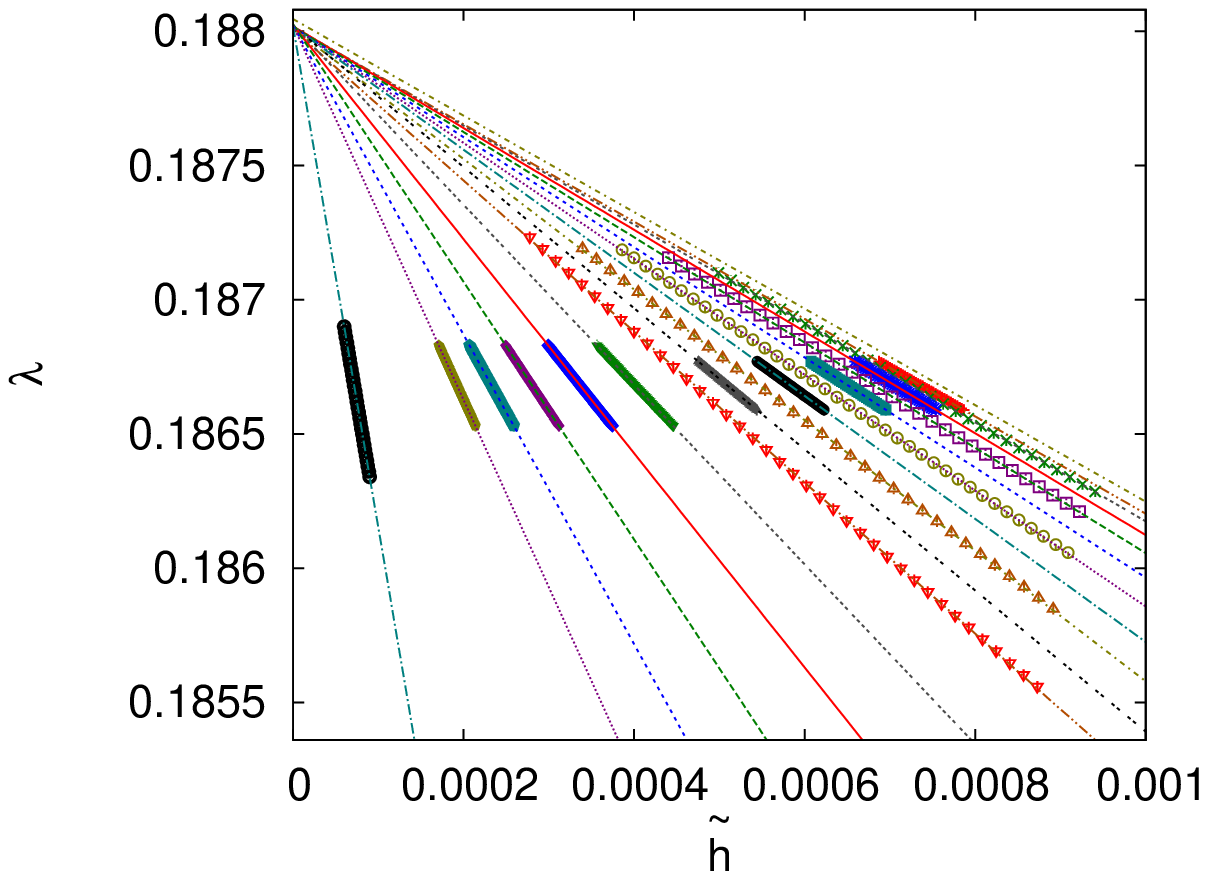} 
}
\caption{Left: Binder cumulant $B_{4,Q}(\hti,\lambda)$ for chemical potential $\mu/T=1.4$ ($N_s = 24$). The thick straight lines mark the 
locus of the minima at $N_s=16,24$, the surrounding contour lines correspond to the values $1.58, 1.64, 1.74$. 
Right: Transition lines $\lambda_{pc}(\hti,\mu/T)$ for several values of $\mu/T$ between $0$ (top) and 
$3.0$ (bottom).}
\label{fig:finitemu_sample}
\end{figure}

For our simulations, we used values of $\mu/T=0.1, \ldots, 3.0$ on three lattice sizes $N_s=16,20,24$. 
Data were produced at a given set $(\lambda,\hti,\mu/T)$ close to the
critical point and later reweighted to near-by values of the couplings.
Over 700 k  configurations were produced for each parameter set and lattice size. 
For each chemical potential, the pseudo-critical line $\lambda_{pc}(\hti)$ was identified as the curve of local minima 
in $B_{4,Q}$, Fig.~\ref{fig:finitemu_sample}, showing again linear behaviour as already observed for $\mu=0$.
Indeed we can repeat the steps leading to Eq.~(\ref{eq:linear_fit}), this time setting 
$f_c(\lambda, \hti,\mu/T) = f_d(\lambda, \hti,\mu/T)$ along the first order line. For small fields $\hti$ we expand about
the Yang-Mills limit $(\lambda, \hti,\mu/T) = (\lambda_0,0,0)$ and obtain
\eq
	\lambda_{pc}(\tilde h, \mu/T) = \lambda_0 + a_1(\mu/T) \tilde h\,, \quad a_1= \frac{2\partial_h f_d}{\partial_\lambda(f_c-f_d)}\bigg|_{(\lambda_0,0,0)}\cosh{(\mu/T)}\;.
	\label{eq:critline_lambda-kappa_mu}
\qe
Fitting to this form ($\chi^2/{\rm dof} \simeq 1.5$) gives
an intercept $\lambda_0 = 0.18802(2)$ consistent with $\lambda_0$ in Eq.~(\ref{eq:linear_fit}) and a slope
\eq
	a_1(\mu/T) = C \cosh(\mu/T)
	\label{eq:a1_mu}
\qe
with $C = -1.814(3)$, which is compatible with Eq.~(\ref{eq:linear_fit}).

\begin{figure}
\centerline{
\includegraphics*[height=0.4\textwidth]{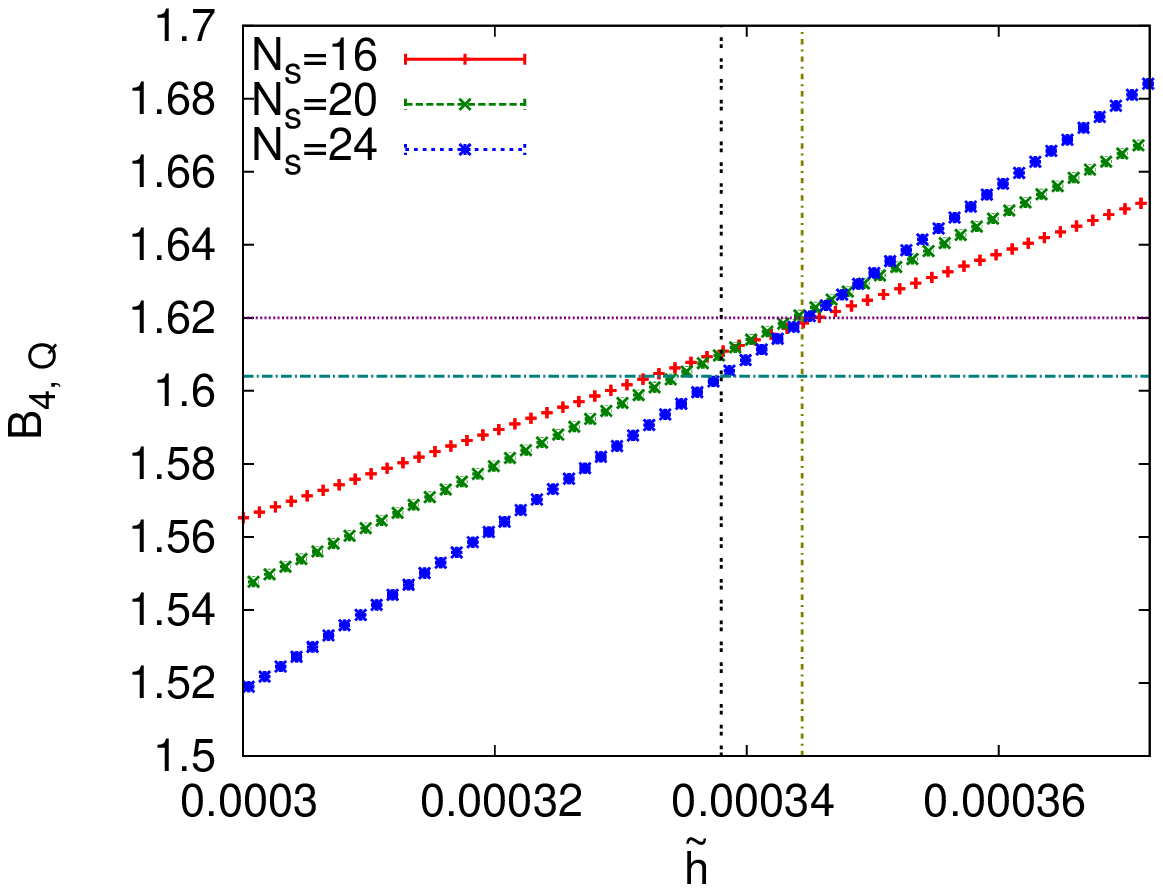}
\includegraphics*[height=0.4\textwidth]{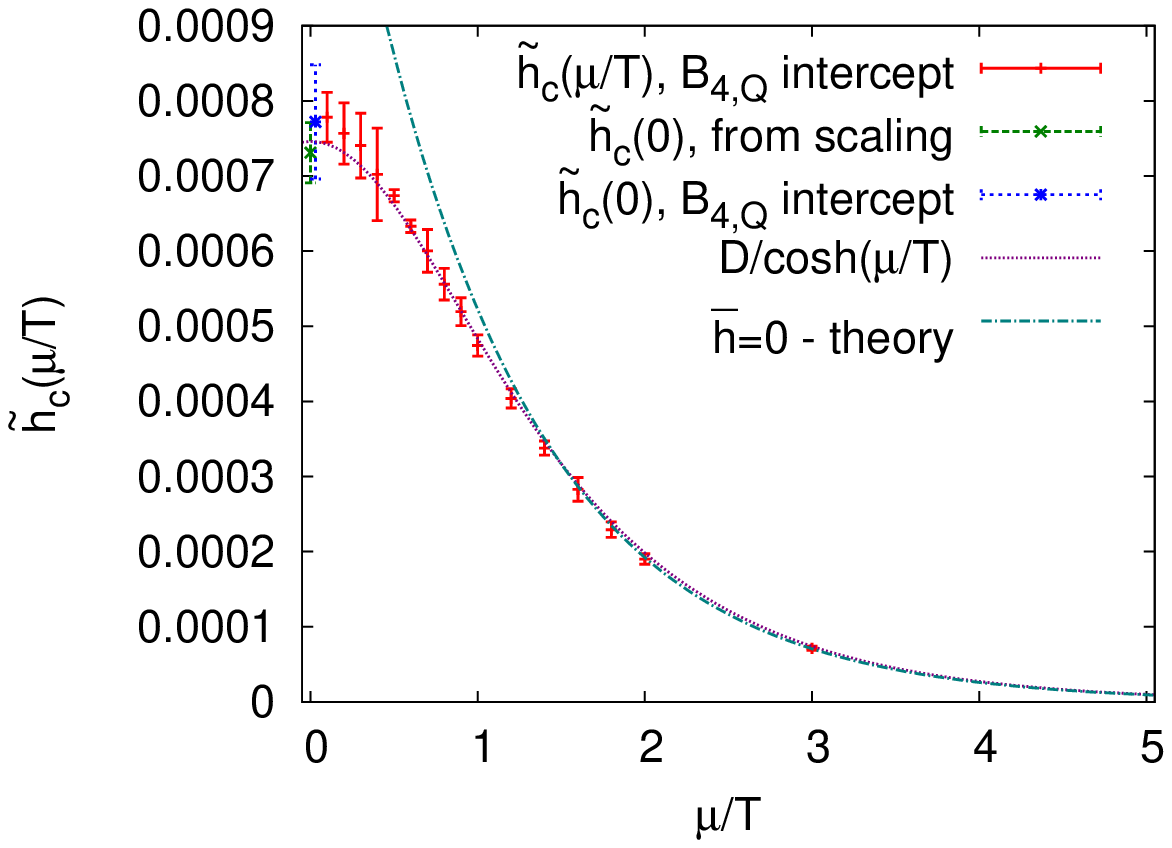} 
}
\caption{Left: $B_{4,Q}$ for different system sizes along the pseudo-critical line for \mbox{$\mu/T = 1.4$}. The horizontal 
lines mark the thermodynamic limit
value of $B_{4,Q}$ at criticality and the actual value of the crossing.
Right:
Critical line $\hti_c(\mu/T)$; the curves are the fit to the $\cosh^{-1}(\mu/T)$ behaviour, Eq.~(\ref{eq:kappa_of_mu}), and the large-$\mu$ asymptotic limit, Eq.~(\ref{eq:largemulim}).
The separate $\mu=0$ determination is compatible with these findings.}
\label{fig:critline_mu_parametrisations}
\end{figure}

As for $\mu=0$, the critical points $\lambda_c(\hti_c(\mu/T),\mu/T)$ can be found again by evaluating 
$B_{4,Q}$ along each $\mu$-line on different volumes. 
To avoid  doing the entire finite size analysis for all parameter sets we take the 
 critical point as the crossing of the $N_s=24$ data with the theoretical value of $1.604$. 
The difference between this procedure and the crossing of individual volumes serves as an estimate for finite size effects.
On the resulting critical line, $\lambda_c(\hti_c(\mu/T),\mu/T)$ shows only weak dependence on $\mu/T$ within our statistical accuracy and varies around $\lambda_c\approx 0.18670(5)$. Note that this 
remains true even for large $\mu/T$, as we shall see in 
Eq.~(\ref{critstat}).\footnote{Interestingly, the same observation is made in the 3d Potts model, where the spin coupling as a function of the external fields is nearly constant along the 
critical line, e.g.~\cite{Kim:2005}.}
We exploit this behaviour to find a simple parametric description of the critical line in terms
of the parameters of the original QCD action. Setting $\Delta \lambda = \lambda_c(\mu/T)-\lambda_0\approx \mathrm{const}$, we may rewrite Eq.~(\ref{eq:critline_lambda-kappa_mu}) at the critical point for fixed $\mu/T$ as
\eq
	\tilde{h}_c (\mu/T)= \frac{\Delta\lambda}{C \cosh (\mu/T)} \equiv \frac{D}{ \cosh (\mu/T)}\;.
	\label{eq:kappa_of_mu}
\qe
A fit of all $\mu>0$ data to Eq.~(\ref{eq:kappa_of_mu}) with only $D$
as parameter performs indeed very well, yielding $D = 0.00075(1)$ with $\chi^2/{\rm dof}=0.6$. The corresponding $h_c(\mu=0) = D$ is also compatible with our findings at $\mu=0$. The resulting curve is shown in Fig.~\ref{fig:critline_mu_parametrisations} (right). While the data seem to be reasonably well described by the Ansatz Eq.~(\ref{eq:kappa_of_mu}), a systematic underestimation at small $\mu/T$ hints towards a more complex law. Indeed our curve is the result of a first order expansion in $(\lambda,h,\bar h)$, Eq.~(\ref{eq:critline_lambda-kappa_mu}), followed by the approximation $\Delta \lambda \approx \mathrm{const}$ and a fit over the whole $\mu$-range. 

On the other hand, asymptotically large chemical potentials in the original lattice QCD
are described by the limit
\eq
	\kappa \to 0, \mu \to \infty\,\mbox{ with} \quad \kappa e^{\mu/T} = \mbox{const}\;.\label{eq:largemulim}
\qe
The corresponding effective theory has two parameters, $(\lambda,h,\hb=0)$. The critical point in this case is easily found by the same techniques, 
\eq
(\lambda_c, h_c)|_{\bar{h}=0}=( 0.18668(2), 0.00142(2))\;.
\label{critstat}
\qe
Using the leading order expression for $h$, Eq.~(\ref{eq:loh}), this gives the $\bar h=0$-critical curve $\hti_{c}(\mu/T)$ which is also plotted in Fig.~\ref{fig:critline_mu_parametrisations}. 
Already for $\mu/T\gsim 1.5$ the data are accurately described by the asymptotic density
limit. Taking this result together with the curve Eq.~(\ref{eq:kappa_of_mu}) we thus have obtained a description of the 
$N_f=1$ deconfinement critical line for all real chemical potentials!

\subsection{Imaginary chemical potential}

The QCD phase transitions and its limiting critical surfaces possess an analytic continuation 
to negative $\mu^2$, or imaginary chemical potentials, $\mu=i\mu_i$. This has been exploited
previously \cite{fp3,fp4}, since in this case the fermion determinant is real positive and properties of the
critical surfaces can be calculated without sign problem. 
In particular, the deconfinement 
critical surface is expected to
terminate in a tricritical line at $\mu_i/T=\pi/3$ \cite{deForcrand:2010he}. 
This value of imaginary chemical potential marks the boundary to an adjacent 
$Z(3)$ centre-sector of the partition function \cite{Roberge:1986mm}.
 Our effective theory correctly reflects the centre symmetry and its breaking in QCD, 
 and hence the related phase structure. 
In this section we explicitly compute the continuation of the critical 
quark masses, i.e.~the deconfinement critical surface, from $\mu=0$ to $\mu/T=i\pi/3$.

Now also the fermionic part of our effective theory, $Q_x$, is explicitly real. Numerically we follow
the same approach as for real $\mu$,  choosing 
values of $\mu_i/T=0.1-\frac{\pi}{3}$, followed by determinations of 
the pseudo-critical and critical couplings.
We observe increasing numerical difficulties as $\mu_i$ approaches
the boundary to the next centre sector. Moving along the critical line towards the Roberge-Weiss tricritical point, a crossover between 3d Ising and tricritical 
scaling sets in, thus obscuring the finite size analysis and demanding ever larger volumes. 
Controlled errors were obtained up to $\mu_i/T \lsim 0.8$.

\begin{figure}
\centerline{
\includegraphics*[height=0.4\textwidth]{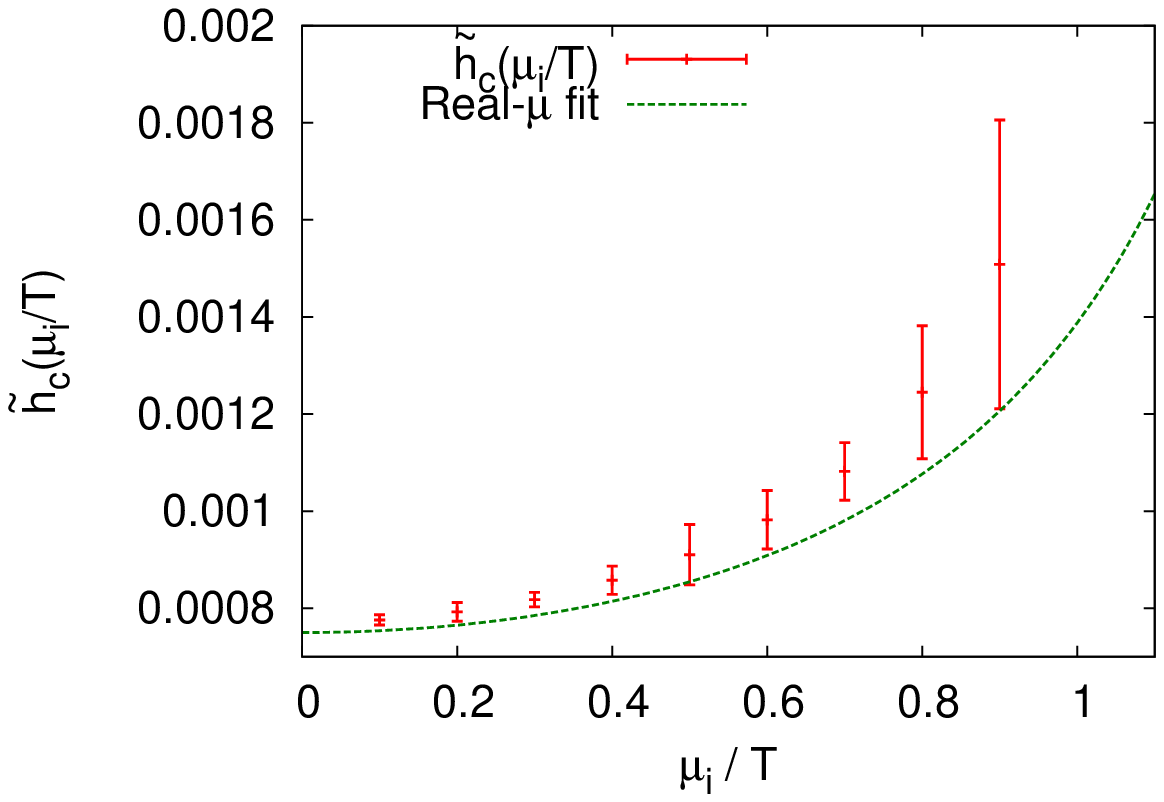} 
\includegraphics[height=0.4\textwidth]{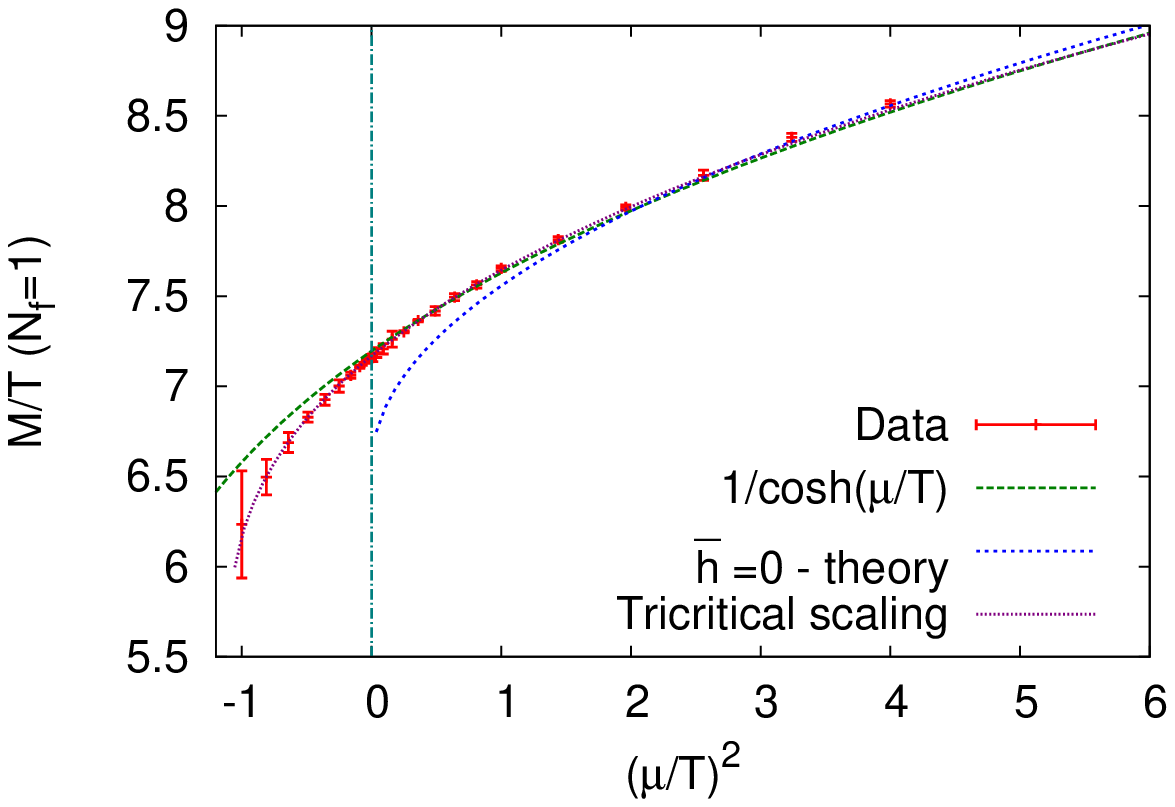}
}
\caption{Left: Critical coupling $\hti_c$ for imaginary chemical potential $\mu_i$, also shown is the analytic continuation of  
the real-$\mu$ fit to Eq.~(\ref{eq:kappa_of_mu}). Right: $M_c/T$ for $N_f=1$ at both imaginary and real chemical potential. The
curves represent Eq.~(\ref{eq:kappa_of_mu}) (and its analytical continuation), the large-$\mu$ asymptote  from Eq.~(\ref{eq:largemulim}), and the tricritical scaling, Eq.~(\ref{eq:tricritical_scaling}).}
\label{fig:imag_critpoint}
\end{figure}

With increasing $\mu_i$, the endpoint of the corresponding first order line is shifted towards higher $\tilde h_c(\mu_i/T)$. The resulting critical line is shown in Fig.~\ref{fig:imag_critpoint} (left). The pseudo-critical lines $\lambda_{pc}(\tilde h,\mu_i/T)$ develop a curvature for increasing $\mu_i$ and thus invalidate a first order expansion of the free energy as done in Eq.~(\ref{eq:critline_lambda-kappa_mu}). As a consequence, an analytic continuation of the real-$\mu$ fit Eq.~(\ref{eq:kappa_of_mu}) to imaginary chemical potential leads to a less satisfying description of the data.
However, we may put real and imaginary $\mu$ data together and plot $M_c/T$ as in
Fig.~\ref{fig:imag_critpoint} (right). (To convert $\hti\rightarrow M$, we use 
Eq.~(\ref{eq:kappa_vs_m-over-t}),
with $h_ce^{-\mu/T}$ instead of simply $h_c$). It was demonstrated in \cite{deForcrand:2010he} for
the Potts model and strong coupling QCD that the critical quark mass at imaginary
chemical potential is governed by tricritical scaling, with a scaling region extending all the way 
to real $\mu$. We thus attempt the corresponding two-parameter fit to
tricritical scaling, 
\eq
	\frac{M_c}{T}\Big(\frac{\mu^2}{T^2}\Big) = \frac{M_{tric}}{T} + K\Big[ \Big(\frac{\pi}{3}\Big)^2 + \Big(\frac{\mu}{T}\Big)^2 \Big]^{2/5}\;,
	\label{eq:tricritical_scaling}
\qe
which is shown in Fig.~\ref{fig:imag_critpoint} (right).
Fitting solely the region $\mu^2<0$ yields $K=1.55(3)$, and $M_{tric}/T= 5.56(3)$ with 
$\chi^2_\mathrm{red}=0.13$. 
Different numbers of flavours can be re-introduced as for $\mu=0$, obtaining 
\eq
	\frac{M_{tric}}{T} = \{ 5.56(3), 6.25(3), 6.66(3) \} \mbox{  , for  } N_f = \{1,2,3\}\;.\label{eq:Mtri}
\qe
Remarkably, the scaling function correctly describes the data up to large
real chemical potentials, in fact well into the region described by asymptotically large $\mu$. 

\subsection{The deconfinement transition for all parameter values}
\begin{figure}
\centerline{
\includegraphics*[scale=0.8]{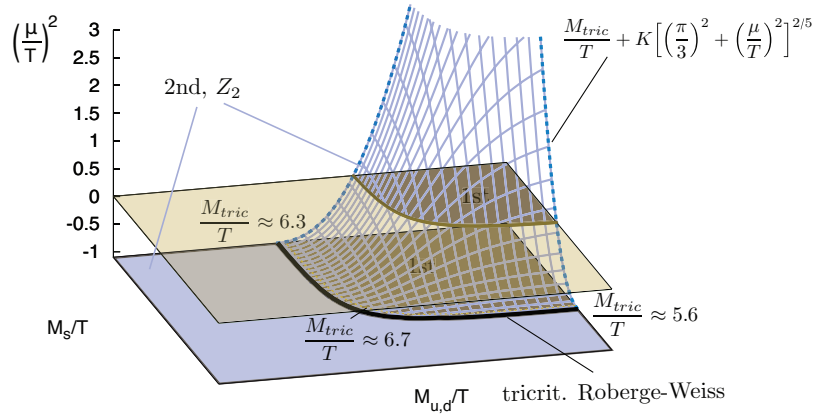} 
}
\caption{The deconfinement critical surface of QCD with heavy quarks.}
\label{fig:columbia_corner_3d}
\end{figure}
Collecting our results from the previous sections,
we can describe the entire deconfinement critical surface from imaginary chemical potentials 
all the way to the large-$\mu$ limit. Moreover, we have a simple and accurate parametrisation
of the surface by stitching together tricritical scaling, the $\cosh^{-1}(\mu/T)$-behaviour for moderate 
$\mu$ and the curve describing the asymptotic limit. 
Converting to a $(2+1)$-flavour setting with different masses $M_{u,d} \neq M_s$,
we may plot the critical surface for the upper-right corner of the Columbia plot (Fig.~\ref{schem}, left).
The final result is shown in Fig.~\ref{fig:columbia_corner_3d}. To convert to quark mass $M/T$ we used Eq.~(\ref{eq:kappa_vs_m-over-t}),
valid to leading order in the hopping expansion and for small $N_\tau \lsim 6$. 

\begin{figure}
\centerline{
\includegraphics*[height=0.35\textwidth]{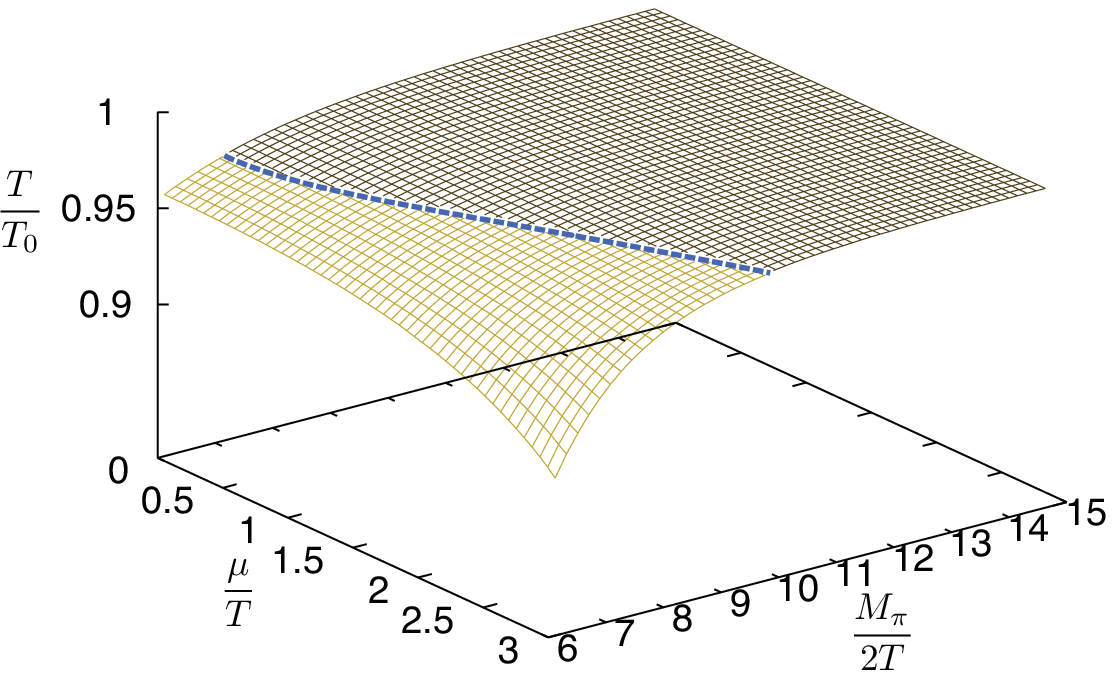}
\includegraphics*[height=0.35\textwidth]{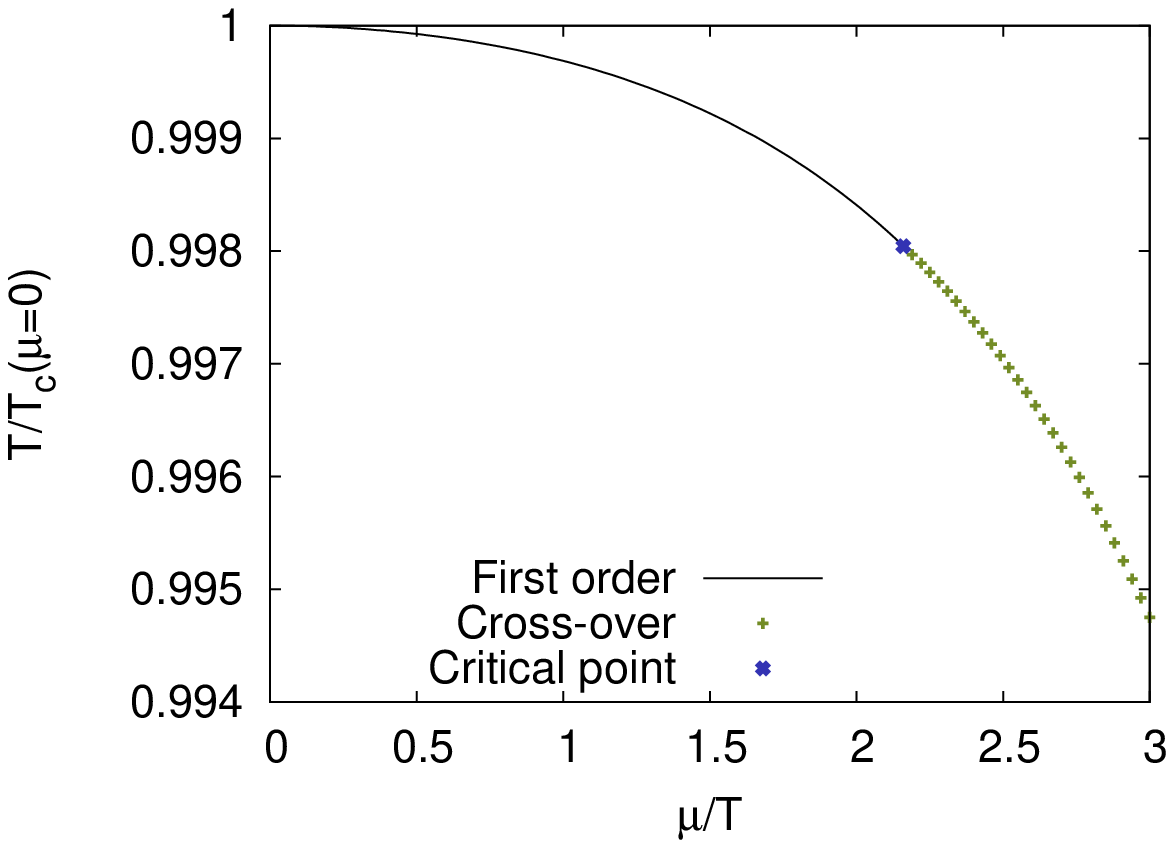} 
}
\caption{Left: Phase structure of the theory in $(\frac{\mu}{T},\frac{M_\pi}{2T}, \frac{T}{T_0})$-space, for  $N_f=2$ and $N_\tau=6$.
Above the surface the theory is deconfined. The critical line (dashed) separates the cross-over (light) and the first-order surface (dark).
Right: Phase diagram of the theory for $M_\pi/(2T)\simeq 8.9$, $N_f=2$ and $N_\tau=6$.}
\label{fig:physical_phase_space}
\end{figure}

By fixing the quark content, e.g.~$N_f=2$, we can similarly draw the full deconfinement transition in  
$(\frac{\mu}{T},\frac{M_\pi}{2T}, \frac{T}{T_0})$-space, with $T_0$ the pure gauge transition temperature.
Since our quarks are very heavy, they give negligible contribution to the beta-function and 
we use again the Sommer parametrisation from pure gauge theory \cite{Sommer} 
to set the scale. As for $\mu=0$, our results are only controlled up to $N_\tau\sim 6$. This yields the phase diagrams in Fig.~\ref{fig:physical_phase_space}. 
The deconfinement temperature for quarks of large but finite mass is almost $\mu$-independent and smoothly approaches a constant value as expected 
for the quenched limit of infinitely heavy quarks.

\section{Conclusions}
\label{sec:con}

We have applied strong coupling and hopping parameter expansions to lattice QCD with Wilson 
fermions at finite temperature and quark chemical potential. The resulting three-dimensional effective theory depends solely on traced Polyakov loops, i.e.~complex scalars, with its couplings given analytically in terms of
the original parameters of the theory, $\beta, \kappa, N_\tau$. 
This leads to a considerable simplification of numerical simulations. In the pure gauge limit we know
five non-trivial orders in the strong coupling expansion, and the resulting
critical parameters are valid up to $N_\tau\sim 16$, enabling a continuum extrapolation of $T_c$.
However, the hopping expansion has only been performed up to order $\kappa^6$, such that our theory with quarks is only robust to $N_\tau\sim 6$ so far. Nevertheless, this corresponds to finer lattices
than have been directly simulated in the heavy quark regime at the time of writing, and moreover offers intriguing features absent in the full 4d theory.
The sign problem is mild enough to directly simulate the model, and can be solved completely within a flux 
representation of the partition function.
Hence a numerical study for all values of the chemical potential is feasible. We have demonstrated this
by computing the entire deconfinement critical surface. While this region of the QCD parameter 
space is far from the physical parameter values, we have presented the first full
QCD lattice calculation involving an arbitrary chemical potential. Furthermore, our results may serve
as benchmarks for analytic approaches, which can easily tune the quark masses.
There are many ways forward from here. As a next step, it would be interesting to study the cold and dense regime within the current effective theory. 
Improvements on the present state can be made by either carrying the analytic calculations to higher order, including additional couplings, or non-perturbatively
by inverse Monte Carlo methods along the lines of  \cite{Wozar:2007tz, deForcrand:2008aw}. Finally and most importantly,
it is most interesting to explore the possibilities of a similar description for the light quark sector, either
by extrapolating a higher-order hopping expansion or by an alternative formulation within the effective
theory context.

\section*{Acknowledgements}
This project is supported by the German BMBF under contract number 06MS9150 and by the Helmholtz International Center for FAIR within the framework of the LOEWE program launched by the State of Hesse.

\end{document}